\pdfoutput=1
\documentclass[final,5p,times, twocolumn]{elsarticle}

\usepackage{amssymb}

\usepackage[labelsep=period,labelfont=bf,justification=raggedright,singlelinecheck=false]{caption}

\usepackage{textcomp}
\usepackage{multirow}
\usepackage{array}
\usepackage{graphicx}
\usepackage{fixltx2e}

\usepackage{url}
\usepackage{mdwlist} 
\usepackage[resetlabels]{multibib}
\newcites{S}{Appendix~~D.~~Selected primary studies}
\newcites{X}{Appendix~~E.~~Typically excluded primary studies}

\journal{Journal of Systems and Software}

\begin{document}

\begin{frontmatter}

\title{On Evaluating Commercial Cloud Services: A Systematic Review}

\author[zli]{Zheng Li}
\author[hzh]{He Zhang}
\author[liam]{Liam O'Brien}
\author[cai]{Rainbow Cai}
\author[liam]{Shayne Flint}
\address[zli]{School of Computer Science, Australian National University \& NICTA, Canberra, Australia}
\address[hzh]{School of Computer Science and Engineering, University of New South Wales, Sydney, Australia}
\address[liam]{School of Computer Science, Australian National University, Canberra, Australia}
\address[cai]{Division of Information, Australian National University, Canberra, Australia}

\begin{abstract}

\textit{Background:} Cloud Computing is increasingly booming in industry with many competing providers and services. Accordingly, evaluation of commercial Cloud services is necessary. However, the existing evaluation studies are relatively chaotic. There exists tremendous confusion and gap between practices and theory about Cloud services evaluation.
\textit{Aim:} To facilitate relieving the aforementioned chaos, this work aims to synthesize the existing evaluation implementations to outline the state-of-the-practice and also identify research opportunities in Cloud services evaluation. 
\textit{Method:} Based on a conceptual evaluation model comprising six steps, the Systematic Literature Review (SLR) method was employed to collect relevant evidence to investigate the Cloud services evaluation step by step. 
\textit{Results:} This SLR identified 82 relevant evaluation studies. The overall data collected from these studies essentially represent the current practical landscape of implementing Cloud services evaluation, and in turn can be reused to facilitate future evaluation work. 
\textit{Conclusions:} Evaluation of commercial Cloud services has become a world-wide research topic. Some of the findings of this SLR identify several research
gaps in the area of Cloud services evaluation (e.g.,~the Elasticity and Security evaluation of commercial Cloud services could be a long-term challenge), while some other findings suggest the trend of applying commercial Cloud services (e.g.,~compared with PaaS, IaaS seems more suitable for customers and is particularly important in industry). This SLR study itself also confirms some previous experiences and reveals new Evidence-Based Software Engineering (EBSE) lessons.  

\end{abstract}


\end{frontmatter}

\section{Introduction}
\label{I}
By allowing customers to access computing services without owning computing infrastructures, Cloud Computing has emerged as one of the most promising computing paradigms in industry \cite{Buyya_Yeo_Venugopal_2009}. Correspondingly, there are more and more commercial Cloud services supplied by an increasing number of providers available in the market \cite{Prodan_Ostermann_2009}\citeS{Li_Yang_2010}.\footnote{We use two types of bibliography formats: the alphabetic format denotes the Cloud service evaluation studies (primary studies) of the SLR, while the numeric format refers to the other references for this article.} Since different and competitive Cloud services may be offered with different terminologies, definitions, and goals \cite{Prodan_Ostermann_2009}, Cloud services evaluation would be crucial and beneficial for both service customers (e.g. cost-benefit analysis) and providers (e.g. direction of improvement) \citeS{Li_Yang_2010}.

However, the evaluation of commercial Cloud services is inevitably challenging for two main reasons. Firstly, previous evaluation results may become quickly out of date. Cloud providers may continually upgrade their hardware and software infrastructures, and new commercial Cloud services and technologies may gradually enter the market. For example, at the time of writing, Amazon
is still acquiring additional sites for Cloud data center expansion \cite{Miller_2011}; Google is moving its App Engine service from CPU usage model to instance model \cite{Alesandre_2011}; while
IBM just offered a public and commercial Cloud \cite{Harris_2011}. As a result, customers
would have to continuously re-design and repeat evaluation for employing
commercial Cloud services.

Secondly, the back-ends (e.g.\ configurations of physical infrastructure) of commercial Cloud services are uncontrollable (often invisible) from the perspective of customers. Unlike consumer-owned computing systems, customers have little knowledge or control over the precise nature of Cloud services even in the \textquotedblleft locked down\textquotedblright { }environment \citeS{Sobel_Subramanyam_2008}. Evaluations in the context of public Cloud Computing are
then inevitably more challenging than that for systems where the customer
is in direct control of all aspects \citeS{Stantchev_2009}. In fact, it is natural that the evaluation
of uncontrollable systems would be more complex than that of controllable ones.

Meanwhile, the existing Cloud services evaluation research is relatively chaotic. On one hand, the Cloud can be viewed from various perspectives \cite{Stokes_2011}, which may result in market hype and also skepticism and confusion \cite{Zhang_Cheng_2010}. As such, it is hard to point out the range of Cloud Computing and a full scope of metrics to evaluate different commercial Cloud services. On the other hand, there exists a tremendous gap between practice and research about Cloud services evaluation. For example, although the traditional benchmarks have been recognized as being insufficient for evaluating commercial Cloud services \cite{Binnig_Kossmann_2009}, they are still predominately used in practice for Cloud services evaluation.

To facilitate relieving the aforementioned research chaos, it is necessary for researchers and practitioners to understand the state-of-the-practice of commercial Cloud services evaluation. For example, the existing evaluation implementations can be viewed as primary evidence for adjusting research directions or summarizing feasible evaluation guidelines. 
As the main methodology applied for Evidence-Based Software Engineering (EBSE) \cite{Dyba_Kitchenham_2005}, 
the Systematic Literature Review (SLR) has been widely accepted as a standard and rigorous approach to evidence aggregation for investigating specific research questions \cite{Kitchenham_Charters_2007,Zhang_Babar_2011}. Naturally, we adopted the SLR method to identify, assess and synthesize the relevant primary studies to investigate Cloud services evaluation. In fact, according to the popular aims of implementing a systematic review \cite{Lisboa_Garcia_2010}, the results of this SLR can help identify gaps in current research and also provide a solid background for future research activities in the field of Cloud services evaluation.  

This paper outlines the work involved in conducting this SLR on evaluating commercial Cloud services. Benefitting from this SLR, we confirm the conceptual model of Cloud services evaluation; the state-of-the-practice of the Cloud services evaluation is finally revealed; and several findings are highlighted as suggestions for future Cloud services evaluation work. In addition to the SLR results, the lessons learned from performing this SLR are also reported in the end. By observing the detailed implementation of this SLR, we confirm some suggestions supplied by the previous SLR studies, and also summarize our own experiences that could be helpful in the community of EBSE \cite{Dyba_Kitchenham_2005}. In particular, to distinguish and elaborate some specific findings, three parts (namely evaluation taxonomy \cite{Li_OBrien_2012a}, metrics \cite{Li_OBrien_2012b}, and factors \cite{Li_OBrien_2012c}) of the outcome derived from this SLR have been reported separately. To avoid duplication, the previously reported results are only briefly summarized (cf.~Subsection \ref{RQ3}, \ref{RQ4}, and \ref{RQ6}) in this paper.

The remainder of this paper is organized as follows. Section \ref{II} supplements the background of this SLR, which introduces a spatial perspective as prerequisite to investigating Cloud services evaluations. Section \ref{III} elaborates the SLR method and procedure employed in this study. Section \ref{IV} briefly describes the SLR results, while Section \ref{V} answers the predefined research questions and highlights the findings. Section \ref{VI} discusses our own experiences in using the SLR method, while Section \ref{VII} shows some limitations with this study. Conclusions and some future work are discussed in Section \ref{VIII}.

\section{Related work and a conceptual model of Cloud services evaluation}
\label{II}
Evaluation of commercial
Cloud services emerged as soon as those services were
published \citeS{Garfinkel_Report_2007,Hill_Li_2010}. In fact, Cloud services evaluation
has rapidly and increasingly become a world-wide research
topic during recent years. As a result, numerous research results have been published, covering various aspects of Cloud services evaluation. Although it is impossible to enumerate all the existing evaluation-related studies, we can roughly distinguish between different studies according to different evaluation aspects on which they mainly focused. Note that, since we are interested in the practices of Cloud services evaluation, \textit{Experiment-Intensive Studies} are the main review objects in this SLR. Based on the rough differentiation, the general process of Cloud services evaluation can be approximately summarized and profiled using a conceptual model.

\subsection{Different studies of Cloud services evaluation}

\textbf{\textit{Service Feature-Emphasized Studies:}}

Since Cloud services are concrete representations of the Cloud Computing paradigm, the Cloud service features to be evaluated have been discussed mainly over Cloud Computing-related introductions, surveys, or research agendas. For example, the characteristics and relationships of Clouds and related technologies were clarified in \cite{Buyya_Yeo_Venugopal_2009,Foster_Zhao_2008,Zhang_Cheng_2010}, which hinted the features that commercial Cloud services may generally embrace. The authors portrayed the landscape of Cloud Computing with regard to trust and reputation \cite{Habib_Ries_2010}. Most of the studies \cite{Armbrust_Fox_2010,Buyya_Yeo_Venugopal_2009,Rimal_Choi_2009,Zhang_Cheng_2010} also summarized and compared detailed features of typical Cloud services in the current market. In particular, the Berkeley view of Cloud Computing \cite{Armbrust_Fox_2010} emphasized the economics when employing Cloud services.\\

\textbf{\textit{Metrics-Emphasized Studies:}}

When evaluating Cloud services, a set of suitable measurement criteria or metrics must be chosen. As such, every single evaluation study inevitably mentions particular metrics when reporting the evaluation process and/or result. However, we did not find any systematic discussion about metrics for evaluating Cloud services. Considering that the selection of metrics plays an essential role in evaluation implementations \cite{Obaidat_Boudriga_2010}, we performed a comprehensive investigation into evaluation metrics in the Cloud Computing domain based on this SLR. The investigation result has been published in \cite{Li_OBrien_2012b}. To the best of our knowledge, this is the only metrics-intensive study of Cloud services evaluation.\\

\textbf{\textit{Benchmark-Emphasized Studies:}}
 
Although traditional benchmarks have been widely employed for evaluating commercial Cloud services, there are concerns that traditional benchmarks may not be sufficient to meet the idiosyncratic characteristics of Cloud Computing. Correspondingly, the authors theoretically portrayed what an ideal Cloud benchmark should be \citeX{Binnig_Kossmann_2009}. In fact, several new Cloud benchmarks have been developed, for example Yahoo!~Cloud Serving Benchmark (YCSB) \citeX{Cooper_Silberstein_2010} and CloudStone \citeS{Sobel_Subramanyam_2008}. In particular, six types of emerging scale-out workloads were collected to construct a benchmark suite, namely CloudSuite \cite{Ferdman_Adileh_2012}, to represent today's dominant Cloud-based applications, such as Data Serving, MapReduce, Media Streaming, SAT Solver, Web Frontend, and Web Search.\\

\textbf{\textit{Experiment-Emphasized Studies:}}

To reveal the rapidly-changing and customer-uncontrollable nature of commercial Cloud services, evaluations have to be implemented through practical experiments. In detail, an evaluation experiment is composed of experimental environment and experimental manipulation. If only focusing on the Cloud side, experimental environment indicates the involved Cloud resources like amount \citeS{Stantchev_2009} or location \citeS{Dejun_Pierre_2010} of service instances, while experimental manipulation refers to the necessary operations on the Cloud resources together with workloads, for example increasing resource amount \citeS{Bientinesi_Iakymchuk_2010} or varying request frequency \citeS{Zhao_Liu_2010}. In fact, given the aforementioned motivation, the existing experiment-intensive studies have been identified and used as the review objects in this SLR.\\

\subsection{A conceptual model of the generic process of Cloud services evaluation}
\begin{figure}
\centering
\includegraphics{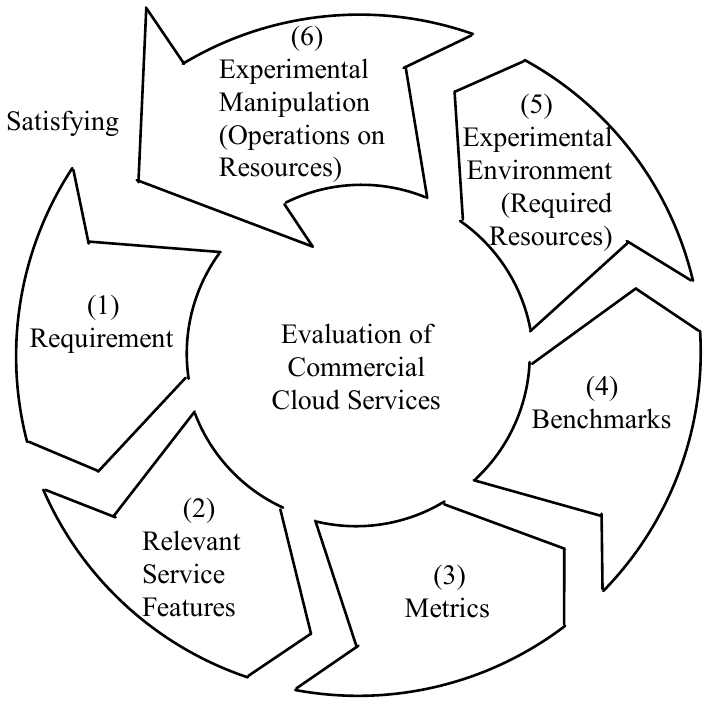}
\caption{\label{fig>1}A conceptual model of the generic process of Cloud services evaluation.}
\end{figure}

As mentioned previously, Cloud Computing is an emerging computing paradigm \cite{Buyya_Yeo_Venugopal_2009}. When it comes to the evaluation of a computing system (commercial Cloud services in this case), one of the most common issues may be the performance evaluation. Therefore, we decided to borrow the existing lessons from performance evaluation of traditional computing systems to investigate the generic process of Cloud services evaluation. In fact, to avoid possible evaluation mistakes, the steps common to all performance evaluation projects have been summarized ranging from \textit{Stating Goals} to \textit{Presenting Results} \cite{Jain_1991}. By adapting these steps to the above-discussed related work, we decomposed an evaluation implementation process into six common steps and built a conceptual model of Cloud services evaluation, as illustrated in Fig.~\ref{fig>1} and specified below.

\begin{enumerate*}
\renewcommand{\labelenumi}{\it{(\theenumi)}}
    \item	First of all, the requirement should be specified to clarify the evaluation purpose, which essentially drives the remaining steps of the evaluation implementation.
    \item	Based on the evaluation requirement, we can identify the relevant Cloud service features to be evaluated.
    \item	To measure the relevant service features, suitable metrics should be determined.
    \item	According to the determined metrics, we can employ corresponding benchmarks that may already exist or have to be developed.
    \item	Before implementing the evaluation experiment, the experimental environment should be constructed. The environment includes not only the Cloud resources to be evaluated but also resources involved in the experiment.
    \item	Given all the aforementioned preparation, the evaluation experiment can be done with human manipulations, which finally satisfies the evaluation requirement.
\end{enumerate*} 

The conceptual model then played a background and foundation role in the conduction of this SLR. Note that this generic evaluation model can be viewed as an abstract of evaluating any computing paradigm. For Cloud services evaluation, the step adaptation is further explained and discussed as a potential validity threat of this study in Subsection \ref{VII>model_threat}.

\section{Review method}
\label{III}
According to the guidelines for performing SLR \cite{Kitchenham_Charters_2007}, we made minor adjustments and planned our study into a protocol. Following the protocol, we unfold this SLR within three stages.
\\

\textbf{\textit{Planning Review:}}
\begin{itemize*}
    \item	Justify the necessity of carrying out this SLR.
    \item	Identify research questions for this SLR.
    \item	Develop SLR protocol by defining search strategy, selection criteria, quality assessment standard, and data extraction schema for Conducting Review stage.
\end{itemize*}

\textbf{\textit{Conducting Review:}}
\begin{itemize*}
    \item	Exhaustively search relevant primary studies in the literature.
    \item	Select relevant primary studies and assess their qualities for answering research questions.
    \item	Extract useful data from the selected primary studies.
    \item	Arrange and synthesize the initial results of our study into review notes.
\end{itemize*}

\textbf{\textit{Reporting Review:}}
\begin{itemize*}
    \item	Analyze and interpret the initial results together with review notes into interpretation notes.
    \item	Finalize and polish the previous notes into an SLR report.
\end{itemize*}

\subsection{Research questions}
\label{III>RQs}
Corresponding to the overall aim of this SLR that is to investigate the procedures and experiences of evaluation of commercial Cloud services, six research questions were determined mainly to address the individual steps of the general evaluation process, as listed in Table \ref{tbl>1}.

\begin{table*}[!t]\footnotesize
\renewcommand{\arraystretch}{1.3}
\centering
\caption{\label{tbl>1}Research questions}
\begin{tabular}{l >{\raggedright}p{5.8cm} >{\raggedright}p{6.4cm} >{\raggedright\arraybackslash}p{3.4cm}}
\hline

\hline
ID & Research Question & Main Motivation & Investigated Step of the General Evaluation Process\\
\hline
RQ1 & What are the purposes of evaluating commercial Cloud services? & To identify the purposes/requirements of evaluating commercial Cloud services. & Requirement\\
RQ2 & What commercial Cloud services have been evaluated? & To identify the most popular Cloud service and its provider that has attracted the dominant research effort. & Service Features\\
RQ3 & What aspects and their properties of commercial Cloud services have been evaluated? & To outline a full scope of aspects and their properties that should be concerned when evaluating Cloud services. & Service Features\\
RQ4 & What metrics have been used for evaluation of commercial Cloud services? & To find metrics practically used in the evaluation of commercial Cloud services. & Metrics\\
RQ5 & What benchmarks have been used for evaluation of commercial Cloud services? & To find benchmarks practically used in the evaluation of commercial Cloud services. & Benchmarks\\
RQ6 & What experimental setup scenes have been adopted for evaluating commercial Cloud services? & To identify the components of environment and operations for building evaluation experiments. & Experimental Environment \& Experimental Manipulation\\
\hline

\hline
\end{tabular}
\end{table*}

In particular, we borrowed the term \textquotedblleft scene\textquotedblright { }from the drama domain for the research question RQ6. In the context of drama, a scene is an individual segment of a plot in a story, and usually settled in a single location. By analogy, here we use \textquotedblleft setup scene\textquotedblright { }to represent an atomic unit for constructing a complete experiment for evaluating commercial Cloud services. Note that, for the convenience of discussion, we broke the investigation of Service Features-oriented step into two research questions (RQ2 and RQ3), while we used one research question (RQ6) to cover both Experimental Environment and Experimental Manipulation steps of the evaluation process (cf.\ Table \ref{tbl>1}).

\subsection{Research scope}
We employed three points in advance to constrain the scope of this research. First, this study focused on the commercial Cloud services only to make our effort closer to industry's needs. Second, this study paid attention to Infrastructure as a Service (IaaS) and Platform as a Service (PaaS) without concerning Software as a Service (SaaS). Since SaaS is not used to further build individual business applications \cite{Binnig_Kossmann_2009}, various SaaS implementations may comprise infinite and exclusive functionalities to be evaluated, which could make this SLR out of control even if adopting extremely strict selection/exclusion criteria. Third, following the past SLR experiences \cite{Sarmad_Ali_2010}, this study also concentrated on the formal reports in academia rather than the informal evaluation practices in other sources.

\subsection{Roles and responsibilities}
\label{III>role}
The members involved in this SLR include a PhD student, a two-people supervisory panel, and a two-people expert panel. The PhD student is new to the Cloud Computing domain, and plans to use this SLR to unfold his research topic. His two supervisors have expertise in the two fields of service computing and evidence-based software engineering respectively, while the expert panel has strong background of computer system evaluation and Cloud Computing. In detail, the expert panel was involved in the discussions about review background, research questions, and data extraction schema when developing the SLR protocol; the specific review process was implemented mainly by the PhD student while under close supervision; the supervisors randomly cross-checked the student's work, for example the selected and excluded publications; regular meetings were held by the supervisory panel with the student to discuss and resolve divergences and confusions over paper selection, data extraction, etc.; unsure issues and data analysis were further discussed by the five members all together.

\subsection{Search strategy and process}
The rigor of the search process is one of the distinctive characteristics of systematic reviews \cite{Zhang_Babar_Tell_2011}. To try to implement an unbiased and strict search, we set a precise publication time span, employed popular literature libraries, alternatively used a set of short search strings, and supplemented a manual search to compensate the automated search for the lack of typical search keywords. 

\subsubsection{Publication time span}
As the term \textquotedblleft Cloud Computing\textquotedblright { }started to gain popularity in 2006 \cite{Zhang_Cheng_2010}, we focused on the literature published from the beginning of 2006. And also considering the possible delay of publishing, we restricted the publication time span between \textbf{January 1st, 2006} and \textbf{December 31st, 2011}.

\subsubsection{Search resources}
\label{III>resources}
With reference to the existing SLR protocols and reports for referential experiences, as well as the statistics of the literature search engines \cite{Zhang_Babar_Tell_2011}, we believed that the following five electronic libraries give a broad enough coverage of relevant primary studies:
\begin{itemize*}
    \item	ACM Digital Library (\url{http://dl.acm.org/})
    \item	Google Scholar (\url{http://scholar.google.com})
    \item	IEEE Xplore (\url{http://ieeexplore.ieee.org})
    \item	ScienceDirect (\url{http://www.sciencedirect.com})
    \item	SpringerLink (\url{http://www.springer.com})
\end{itemize*}

\subsubsection{Proposing search string}
\label{III>string}
We used a three-step approach to proposing search string for this SLR:

\begin{enumerate*}
\renewcommand{\labelenumi}{\it{(\theenumi)}}
    \item	Based on the keywords and their synonyms in the research questions, we first extracted potential search terms, such as:\textgravedbl cloud computing\textacutedbl , \textgravedbl cloud provider\textacutedbl , \textgravedbl cloud service\textacutedbl , evaluation, benchmark, metric, etc.
    \item	Then, by rationally modifying and combining these search terms, we constructed a set of candidate search strings.
    \item	At last, following the Quasi-Gold Standard (QGS) based systematic search approach \cite{Zhang_Babar_Tell_2011}, we performed several pilot manual searches to determine the most suitable search string according to the search performance in terms of sensitivity and precision.
\end{enumerate*} 

Particularly, the sensitivity and precision of a search string can be calculated as shown in Equation (\ref{equation>1}) and (\ref{equation>2}) respectively \cite{Zhang_Babar_Tell_2011}.

\begin{small}
\begin{equation}
\label{equation>1}
Sensitivity=\frac{Number~of~relevant~studies~retrieved}{Total~number~of~relevant~studies}100\%
\end{equation}
\end{small}

\begin{small}
\begin{equation}
\label{equation>2}
Precision=\frac{Number~of~relevant~studies~retrieved}{Number~of~studies~retrieved}100\%
\end{equation}
\end{small}

\begin{table}[!t]\footnotesize
\renewcommand{\arraystretch}{1.3}
\centering
\caption{\label{tbl>2}Sensitivity and precision of the search string with respect to several conference proceedings.}
\begin{tabular}{l l l}
\hline

\hline
Target Proceedings & Sensitivity & Precision\\
\hline
CCGRID 2009 & 100\% (1/1)& 100\% (1/1) \\
CCGRID 2010 & N/A (0/0)& N/A (0/2)\\
CCGRID 2011 & 100\% (1/1) & 50\% (1/2)\\
CloudCom 2010 & 100\% (3/3)& 27.3\% (3/11) \\
CloudCom 2011 & 100\% (2/2) & 33.3\% (2/6)\\
CLOUD 2009 & N/A (0/0)& N/A (0/0)\\
CLOUD 2010 & N/A (0/0)& N/A (0/6)\\
CLOUD 2011 & 66.7\% (2/3) & 25\% (2/8)\\
GRID 2009 & 100\% (1/1)& 50\% (1/2) \\
GRID 2010 & 100\% (1/1)& 100\% (1/1) \\
GRID 2011 & N/A (0/0) & N/A (0/0)\\
\hline
Total & 91.7\% (11/12) & 28.2\% (11/39) \\
\hline

\hline
\end{tabular}
\end{table}

In detail, we selected seven Cloud-related conference proceedings (cf.~Table \ref{tbl>2}) to test and contrast sensitivity and precision of different candidate search strings. According to the suggestions of search strategy scales \cite{Zhang_Babar_Tell_2011}, we finally proposed a search string with the \textit{Optimum} strategy, as shown below:\\
\\
\textbf{(\textgravedbl cloud computing\textacutedbl { }OR \textgravedbl cloud platform\textacutedbl { }OR \textgravedbl cloud provider\textacutedbl { }OR \textgravedbl cloud service\textacutedbl { }OR \textgravedbl cloud offering\textacutedbl ) AND (evaluation OR evaluating OR evaluate OR evaluated OR experiment OR benchmark OR metric OR simulation) AND (\textless Cloud provider's name\textgreater ~OR ...)}\\

Note that the \textbf{(\textless Cloud provider's name\textgreater ~OR ...)} denotes the ``OR"-connected names of the top ten Cloud providers \cite{SearchCloudComputing_2010}. The specific sensitivity and precision of this search string with respect to those seven proceedings are listed in Table \ref{tbl>2}. Given such high sensitivity and more than enough precision \cite{Zhang_Babar_Tell_2011}, although the search string was locally optimized, we have more confidence to expect a globally acceptable search result.

\subsubsection{Study identification process}
\label{III>process}

\begin{figure}[!t]
\centering
\includegraphics{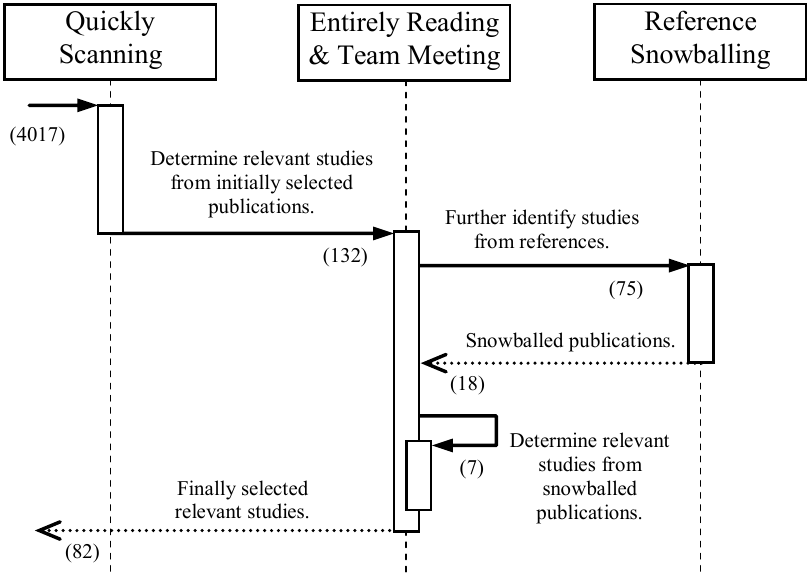}
\caption{\label{fig>PicSequenceDiagram}Study identification process in sequence diagram. The numbers in the brackets denote how many publications were identified/selected at different steps.}
\end{figure}

There are three main activities in the study identification process, as listed below: Quickly Scanning based on the automated search, Entirely Reading and Team Meeting for the initially identified studies, and manual Reference Snowballing. The whole process of study identification has been illustrated as a sequence diagram in Fig.~\ref{fig>PicSequenceDiagram}.\\

\textbf{\textit{(1) Quickly Scanning:}} 

Given the pre-determined search strings, we unfolded automated search in the aforementioned electronic libraries respectively. Relevant primary studies were initially selected by scanning titles, keywords and abstracts. \\

\textbf{\textit{(2) Entirely Reading and Team Meeting:}} 

The initially identified publications were decided by further reviewing the full-text, while the unsure ones were discussed in the team meeting.\\

\textbf{\textit{(3) Reference Snowballing:}} 

To further find possibly missed publications, we also supplemented a manual search by snowballing the references \cite{Kitchenham_Li_2011} of the selected papers found by the automated search. The new papers identified by reference snowballing were also read thoroughly and/or disscussed.

\subsection{Inclusion and exclusion criteria}
In detail, the inclusion and exclusion criteria can be specified as:\\
\\
\textbf{\textit{Inclusion Criteria:}}
\begin{enumerate*}
\renewcommand{\labelenumi}{\it{(\theenumi)}}
    \item	Publications that describe practical evaluation of commercial Cloud services.
    \item	Publications that describe evaluation tool/method/framework for Cloud Computing, and include practical evaluation of commercial Cloud services as a demonstration or case study.
    \item	Publications that describe practical evaluation of comparison or collaboration between different computing paradigms involving commercial Cloud services.
    \item	Publications that describe case studies of adapting or deploying the existing applications or systems to public Cloud platforms with evaluations. This scenario can be viewed as using real applications to benchmark commercial Cloud services. Note the difference between this criterion and Exclusion Criterion (3).
    \item	In particular, above inclusion criteria apply only to regular academic publications (Full journal / conference / workshop papers, technical reports, and book chapters).
\end{enumerate*}
\textbf{\textit{Exclusion Criteria:}}
\begin{enumerate*}
\renewcommand{\labelenumi}{\it{(\theenumi)}}
    \item	Publications that describe evaluation of non-commercial Cloud services in the private Cloud or open-source Cloud.
    \item	Publications that describe only theoretical (non-practical) discussions, like \citeX{Binnig_Kossmann_2009} (cf.~Table \ref{tbl>3aa}), about evaluation for adopting Cloud Computing.
    \item	Publications that propose new Cloud-based applications or systems, and the aim of the corresponding evaluation is merely to reflect the performance or other features of the proposed application/system. Note the difference between this criterion and Inclusion Criterion (4).
    \item	Publications that are previous versions of the later published work.
    \item	In addition, short/position papers, demo or industry publications are all excluded.
\end{enumerate*}

\subsection{Quality assessment criteria}
Since a relevant study can be assessed only through its report, and Cloud services evaluation belongs to the field of experimental computer science \citeS{Stantchev_2009}, here we followed the reporting structure of experimental studies (cf.~Table 9 in \cite{Runeson_2009}) to assess the reporting quality of one publication. In particular, we divided the reporting structural concerns into two categories: the generic Research Reporting quality and the experimental Evaluation Reporting quality.
\begin{itemize*}
    \item	Research Reporting: Is the paper or report well organized and presented following a regular research procedure?
    \item	Evaluation Reporting: Is the evaluation implementation work described thoroughly and appropriately?
\end{itemize*}
In detail, we proposed eight criteria as a checklist to examine different reporting concerns in a relevant study:\\
\\
\textbf{\textit{Criteria of Research Reporting Quality:}}
\begin{enumerate*}
\renewcommand{\labelenumi}{\it{(\theenumi)}}
    \item	Is the research problem clearly specified?
    \item	Are the research aim(s)/objective(s) clearly identified?
    \item	Is the related work comprehensively reviewed?
    \item	Are findings/results reported?
\end{enumerate*}
\textbf{\textit{Criteria of Evaluation Reporting Quality:}}
\begin{enumerate*}
\renewcommand{\labelenumi}{\it{(\theenumi)}}
\setcounter{enumi}{4}
    \item	Is the period of evaluation work specified?
    \item	Is the evaluation environment clearly described?
    \item	Is the evaluation approach clearly described?
    \item	Is the evaluation result analyzed or discussed?
\end{enumerate*}

Each criterion was used to judge one aspect of the quality of a publication, and to assign a quality score for the corresponding aspect of the publication. The quality score can be 1, 0.5, or 0, which represent the quality from excellent to poor as answering Yes, Partial, or No respectively. The overall quality of a publication can then be calculated by summing up all the quality scores received.

\subsection{Data extraction and analysis}
\label{III>schema}
According to the research questions we previously identified, this SLR used a data extraction schema to collect relevant data from primary studies, as listed in Table \ref{tbl>extraction}. The schema covers a set of attributes, and each attribute corresponds to a data extraction question. The relationships between the data extraction questions and predefined research questions are also specified. 

In particular, the collected data can be distinguished between the metadata of publications and experimental data of evaluation work. The metadata was mainly used to perform statistical investigation of relevant publications, while the Cloud services evaluation data was analyzed to answer those predefined research questions. Moreover, the data of \textit{evaluation time} collected by question (14) was used in the quality assessment; the data extraction question (15) about detailed \textit{configuration} was to snapshot the evaluation experiments for possible replication of review. 

\begin{table*}[!t]\footnotesize
\renewcommand{\arraystretch}{1.3}
\centering
\caption{\label{tbl>extraction}The data extraction schema.}

\begin{tabular}{l >{\raggedright}p{1.9cm} >{\raggedright}p{7.5cm} >{\raggedright}p{2.5cm} >{\raggedright\arraybackslash}p{3.2cm} }
\hline

\hline
ID & Data Extraction Attribute & Data Extraction Question & Corresponding Research Question & Investigated Step in the General Evaluation Process\\
\hline
(1) & Author & Who is/are the author(s)?& \multirow{8}{*}{N/A (Metadata)} & \multirow{8}{3.2cm}{N/A (Generic investigation in SLR)}\\
(2) & Affiliation & What is/are the authors' affiliation(s)? &  & \\
(3) & Publication title & What is the title of the publication? & & \\
(4) & Publication year  & In which year was the evaluation work published? & &\\
(5) & Venue type  & What type of the venue does the publication have? (Journal, Conference, Workshop, Book Chapter, or Technical Report) & &\\
(6) & Venue name  & Where is the publication's venue? (Acronym of name of journal, conference, workshop, or institute, e.g.\ ICSE, TSE) & &\\
\hline
(7) & Purpose & What is the purpose of the evaluation work in this study? & RQ1 & Requirement\\
\hline
(8) & Provider & By which commercial Cloud provider(s) are the evaluated services supplied? & \multirow{3}{*}{RQ2} & \multirow{3}{*}{Service Features}\\
(9) & Service  & What commercial Cloud services were evaluated? &&\\
\hline
(10) & Service aspect & What aspect(s) of the commercial Cloud services was/were evaluated in this study?& \multirow{3}{*}{RQ3} & \multirow{3}{*}{Service Features}\\
(11) & Aspect property  & What properties were concerned for the evaluated aspect(s)? &&\\
\hline
(12) & Metric  & What evaluation metrics were used in this study? & RQ4 & Metrics\\
\hline
(13) & Benchmark  & What evaluation benchmark(s) was/were used in this study? & RQ5 & Benchmarks\\
\hline
(14) & Environment  & What environmental setup scene(s) were concerned in this study? & \multirow{2}{*}{RQ6} & Experimental Environment\\
(15) & Operation & What operational setup scene(s) were concerned in this study? & & Experimental Manipulation\\
\hline
(16) & Evaluation time & If specified, when was the time or period of the evaluation work? & N/A (Additional data) & N/A (To note evaluation time/period)\\
\hline
(17) & Configuration  & What detailed configuration(s) was/were made in this study? & N/A (Additional data) & N/A (To facilitate possible replication of review)\\
\hline

\hline
\end{tabular}

\end{table*}

\section{Review results}
\label{IV}
To distinguish the metadata analysis from the evaluation data analysis in this SLR, we first summarize the results of metadata analysis and quality assessment in this section. The findings and answers to those predefined research questions are then discussed in the next section.

Following the search sequence (cf.~Fig.~\ref{fig>PicSequenceDiagram}), 82 relevant primary studies in total were identified. In detail, the proposed search string initially brought 1198, 917, 225, 366 and 1281 results from the ACM Digital Library, Google Scholar, IEEE Xplore, ScienceDirect, and SpringerLink respectively, as listed in the column \textit{Number of Retrieved Papers} of Table \ref{tbl>4}.

By reading titles and abstracts, and quickly scanning publications in the automated search process, we initially gathered 132 papers. After entirely reading these papers, 75 were selected for this SLR. In particular, 17 undecided papers were finally excluded after our discussion in team meetings; two technical reports and four conference papers were excluded due to the duplication of their latter versions. A set of typical excluded papers (cf.~Appendix E) were particularly explained to demonstrate the application of predefined exclusion criteria, as shown in \ref{Appendix>III}. Finally, seven more papers were chosen by reference snowballing in the manual search process. The finally selected 82 primary studies have been listed in Appendix D. The distribution of the identified publications from different electronic databases is listed in Table \ref{tbl>4}. Note that the four manually-identified papers were further located by using Google Scholar.

\begin{figure}[!t]
\centering
\includegraphics[width=8.5cm]{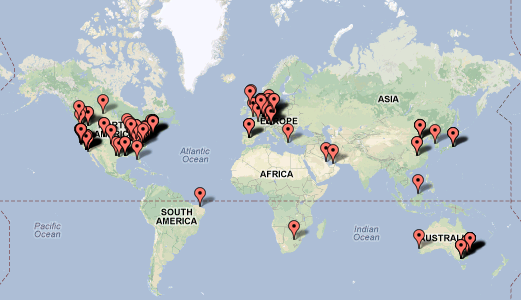}
\caption{\label{fig>2}Study distribution over the (co-)author's affiliations.}
\end{figure}

\begin{table}[!t]\footnotesize
\renewcommand{\arraystretch}{1.3}
\centering
\caption{\label{tbl>4}Distribution of relevant studies over electronic libraries.}
\begin{tabular}{l >{\raggedright}p{1.3cm} >{\raggedright}p{1.3cm} >{\raggedright\arraybackslash}p{1.7cm}}
\hline

\hline
Electronic Library & Number of Retrieved Papers & Number of Relevant Papers & Percentage in Total Relevant Papers\\
\hline
ACM Digital Library & 1198 & 21 & 25.6\%\\
Google Scholar & 917 & 14 & 17.1\%\\
IEEE Xplore & 255 & 36 & 43.9\%\\
ScienceDirect & 366&  0 & 0\%\\
SpringerLink & 1281 & 11 & 13.4\%\\
\hline
Total & 4017 & 82 & 100\%\\
\hline

\hline
\end{tabular}
\end{table}
These 82 primary studies were conducted by 244 authors (co-authors) in total. 40 authors were involved in more than one evaluation works. Interestingly, only four primary studies included co-authors with a direct affiliation with a Cloud services vendor (i.e.~Microsoft). On one hand, it may be fairer and more acceptable for third parties' evaluation work to be published. On the other hand, this phenomenon may result from the limitation with our research scope (cf.~Subsection \ref{VII>scope_threat}). To visibly illustrate the distribution of authors' affiliations, we mark their locations on a map, as shown in Fig.\ \ref{fig>2}. Note that the amount of authors' affiliations is more than the total number of the selected primary studies, because some evaluation work could be collaborated between different research organizations or universities. The map shows that, although major research efforts were from USA, the topic of evaluation of commercial Cloud services has been world-widely researched.

\begin{figure}[!t]
\centering
\includegraphics[width=8.5cm]{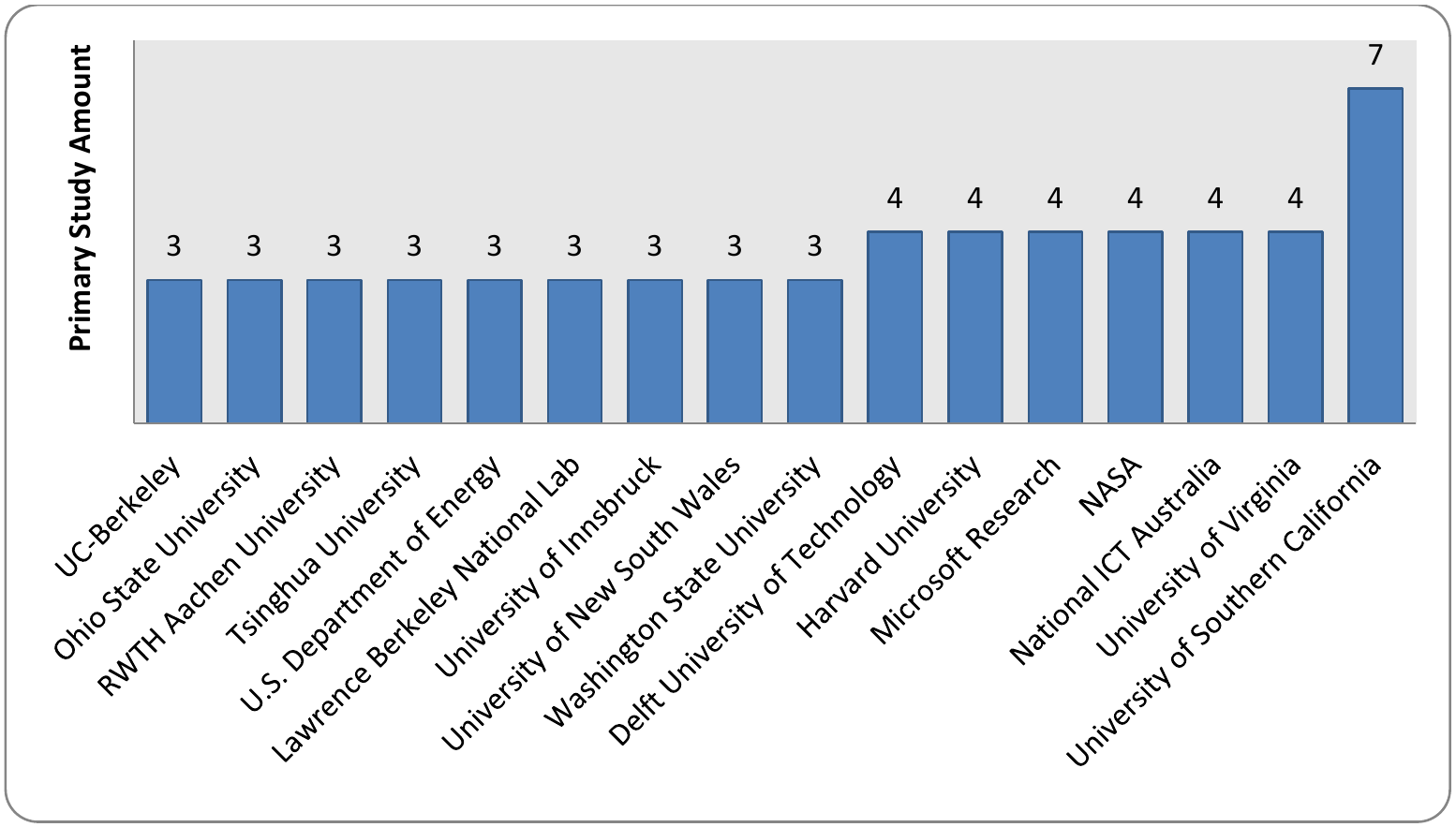}
\caption{\label{fig>affiliations}Universities/Organizations with three or more publications.}
\end{figure}

Furthermore, we can make those affiliations be accurate to: (1) the background universities of institutes, departments or schools; and (2) the background organizations of individual research laboratories or centers. In this paper, we only focus on the universities/organizations that have published three or more primary studies, as shown in Fig.\ \ref{fig>affiliations}. We believe these universities/organizations may have more potential to provide further and continual work on evaluation for commercial Cloud services in the future.

\begin{figure}[!t]
\centering
\includegraphics[width=8.5cm]{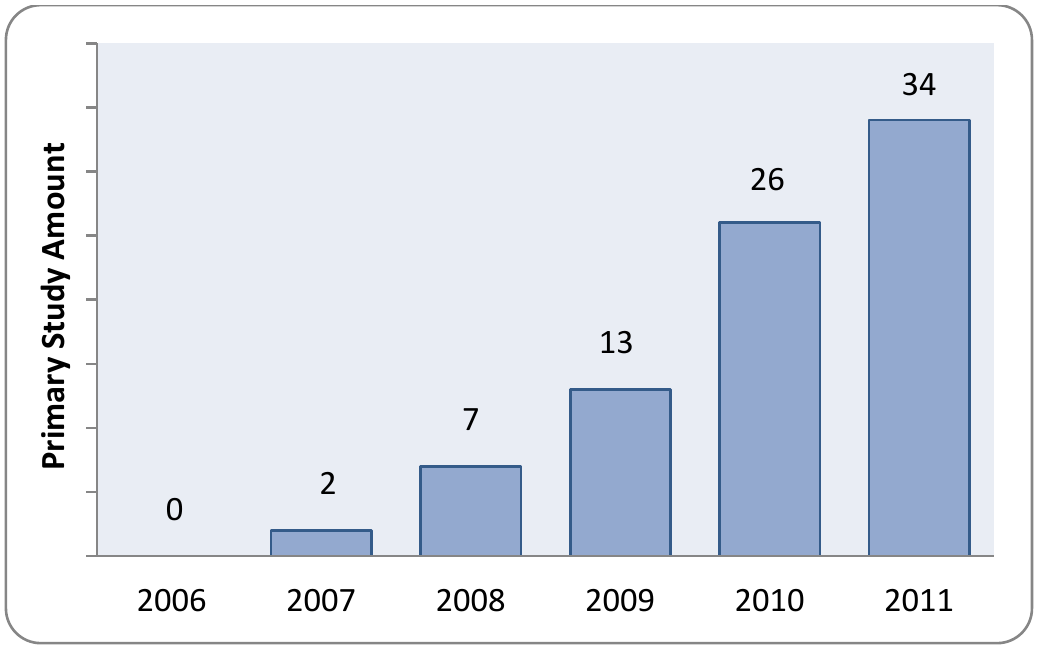}
\caption{\label{fig>year}Study distribution over the publication years.}
\end{figure}

The distribution of publishing time can be illustrated by grouping the primary studies into years, as shown in Fig.\ \ref{fig>year}. It is clear that the research interests in evaluation of commercial Cloud services have been rapidly increased during the past five years.

\begin{figure}[!t]
\centering
\includegraphics[width=7.5cm]{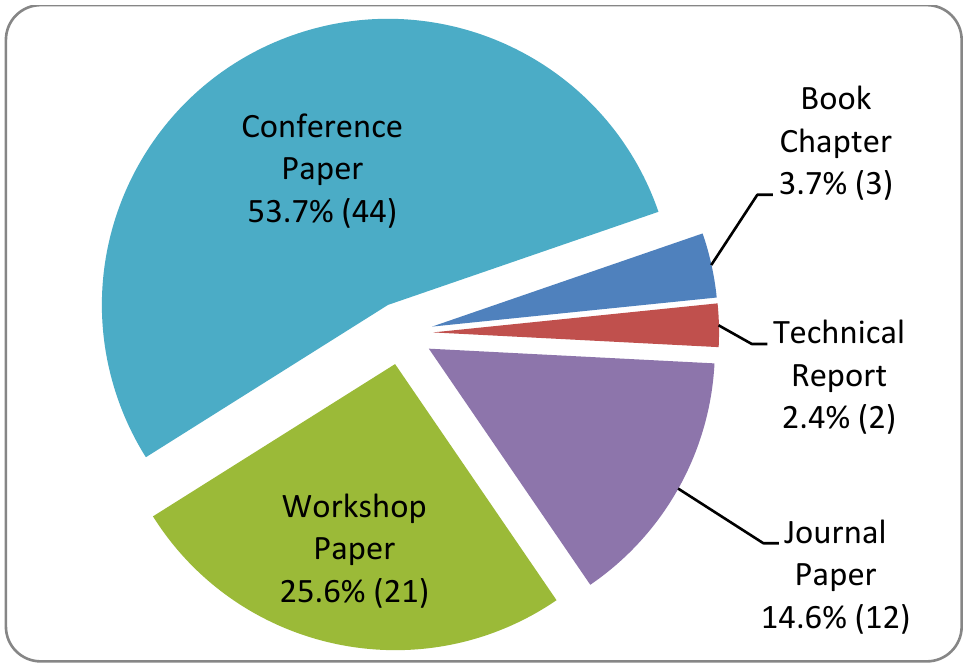}
\caption{\label{fig>venues}Study distribution over the publishing venue types.}
\end{figure}

In addition, these 82 studies on evaluation of commercial Cloud services scattered in as many as 57 different venues. Such a number of publishing venues are more dispersive than we expected. Although there was not a dense publication zone, in general, those venues could be categorized into five different types: Book Chapter, Technical Report, Journal, Workshop, and Conference, as shown in Fig.\ \ref{fig>venues}. Not surprisingly, the publications of evaluation work were relatively concentrated in the Cloud and Distributed Computing related conferences, such as CCGrid, CloudCom, and IPDPS. Moreover, the emerging and Cloud-dedicated books, technical reports, and workshops were also typical publishing venues for Cloud services evaluation work.

As for the quality assessment, instead of listing the detailed quality scores in this paper, here we only show the distribution of the studies over their total reporting quality and total working quality respectively, as listed in Table \ref{tbl>5}.
\begin{table}[!t]\footnotesize
\renewcommand{\arraystretch}{1.3}
\centering
\caption{\label{tbl>5}Distribution of studies over quality.}
\begin{tabular}{p{1.75cm} l l l}
\hline

\hline
Type & Score & Number of Papers & Percentage\\
\hline
\multirow{6}{1.75cm}{Research Reporting Quality} & 2 & 2 & 2.44\%\\
& 2.5 & 2 & 2.44\%\\
& 3 & 22 & 26.83\%\\
& 3.5 & 3 & 3.66\%\\
& 4 & 53 & 64.63\%\\
\cline{2-4}
& Total & 82 & 100\%\\
\hline
\multirow{7}{1.75cm}{Evaluation Reporting Quality} & 1 & 1 & 1.22\%\\
& 2 & 8 & 9.76\%\\
& 2.5 & 13 & 15.85\%\\
& 3 & 17 & 45.12\%\\
& 3.5 & 13 & 15.85\%\\
& 4 & 10 & 12.2\%\\
\cline{2-4}
& Total & 82 & 100\%\\
\hline

\hline
\end{tabular}
\end{table}

According to the quality assessment, in particular, we can highlight two limitations of the existing Cloud services evaluation work. Firstly, less than $16\%$ publications specifically recorded the time of evaluation experiments. As mentioned earlier, since commercial Cloud services are rapidly changing, the lack of exposing experimental time would inevitably spoil reusing evaluation results or tracking past data in the future. Secondly, some primary studies did not thoroughly specify the evaluation environments or experimental procedures. As a result, it would be hard for others to replicate the evaluation experiments or learn from the evaluation experiences reported in those studies, especially when their evaluation results became out of date.

\section{Discussion addressing research questions}
\label{V}
The discussion in this section is naturally organized following the sequence of answers to the six predefined research questions.
\subsection{RQ 1: What are the purposes of evaluating commercial Cloud services?}
\begin{figure}[!t]
\centering
\includegraphics[width=6.5cm]{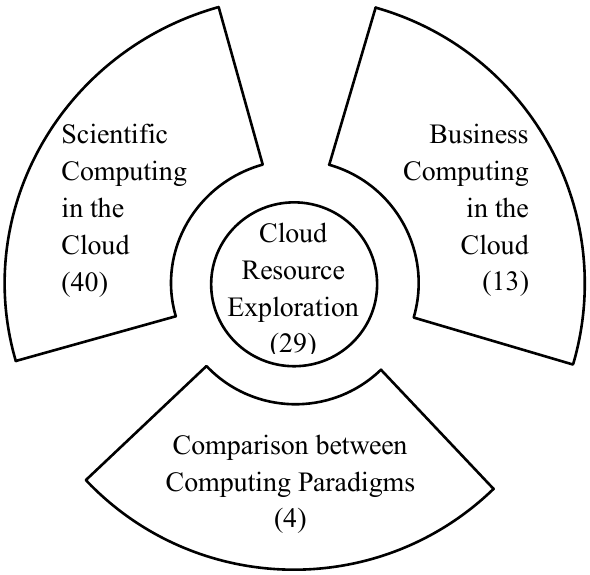}
\caption{\label{fig>5}Purposes of Cloud services evaluation.}
\end{figure}

After reviewing the selected publications, we have found mainly four different motivations behind the evaluations of commercial Cloud services, as illustrated in Fig.\ \ref{fig>5}.   

The \textit{Cloud Resource Exploration} can be viewed as a root motivation. As the name suggests, it is to investigate the available resources like computation capability supplied by commercial Cloud services. For example, the purpose of study \citeS{Stantchev_2009} was to purely understand the computation performance of Amazon EC2. The other three research motivations are essentially consistent with the \textit{Cloud Resource Exploration}, while they have specific intentions of applying Cloud resources, i.e., \textit{Scientific/Business Computing in the Cloud} is to investigate applying Cloud Computing to Scientific/Business issues, and \textit{Comparison between Computing Paradigms} is to compare Cloud Computing with other computing paradigms. For example, study \citeS{Jackson_Ramakrishnan_2010} particularly investigated high-performance scientific computing using Amazon Web services; the benchmark Cloudstone \citeS{Sobel_Subramanyam_2008} was proposed to evaluate the capability of Cloud for hosting Web 2.0 applications; the study \citeS{Carlyle_Harrell_2010} performed a contrast between Cloud Computing and Community Computing with respect to cost effectiveness.
\begin{table}[!t]\footnotesize
\renewcommand{\arraystretch}{1.3}
\centering
\caption{\label{tbl>6}Distribution of studies over evaluation purpose.}
\begin{tabular}{>{\raggedright}p{2.54cm} p{5cm}}
\hline

\hline
Purpose & Primary Studies\\
\hline
Cloud Resource Exploration & \citeS{Alhamad_Dillon_2010} \cite{Bicer_Chiu_2011} \citeS{Baun_Kunze_2009} \citeS{Brebner_Liu_2010} \citeS{Bermbach_Tai_2011} \citeS{Chiu_Agrawal_2010} \citeS{Chen_Bai_2011} \citeS{Chiu_Hall_2011} \citeS{Assuncao_Costanzo_2010} \citeS{Ghoshal_Canon_2011} \citeS{Gent_Kotthoff_2011} \citeS{Garfinkel_Report_2007} \citeS{Hill_Li_2010} \citeS{Islam_Lee_2011} \citeS{Iosup_Yigitbasi_2011} \citeS{Li_Yang_2010} \citeS{Liu_Wee_2009} \citeS{Phillips_Engen_2011} \citeS{Ramasahayam_Deters_2011} \citeS{Ristenpart_Tromer_2009} \citeS{Schad_Dittrich_2010} \citeS{Stantchev_2009} \citeS{Salah_Saba_2011} \citeS{Toyoshima_Yamaguchi_2010} \citeS{Virvilis_Dritsas_2011} \citeS{Wang_Ng_2010} \citeS{Wang_Varman_2011} \citeS{Yigitbasi_Iosup_2009} \citeS{Zhao_Liu_2010}\\

Business Computing in the Cloud & \citeS{Barker_Shenoy_2010} \citeS{Brantner_Florescu_2008} \citeS{Cunha_Mendonca_2011} \citeS{Cervino_Rodriguez_2011} \citeS{Dejun_Pierre_2010} \citeS{Gropengie_Baumann_2011} \citeS{Garfinkel_Journal_2007} \citeS{Gropengie_Sattler_2011} \citeS{Jayasinghe_Malkowski_2011} \citeS{Kossmann_Kraska_2010} \citeS{Lenk_Menzel_2011} \citeS{Li_Yang_2010} \citeS{Sobel_Subramanyam_2008}\\

Scientific Computing in the Cloud & \citeS{Akioka_Muraoka_2010} \citeS{Bientinesi_Iakymchuk_2010} \citeS{Dalman_Doernemann_2010} \citeS{Deelman_Singh_2008} \citeS{Evangelinos_Hill_2008} \citeS{Gunarathne_Wu_2011} \citeS{Hazelhurst_2008} \citeS{Hill_Humphrey_2009} \citeS{Humphrey_Hill_2011} \citeS{He_Zhou_2010} \citeS{Iakymchuk_Napper_2011} \citeS{Iosup_Ostermann_2011} \citeS{Juve_Deelman_2009} \citeS{Juve_Deelman_2010} \citeS{Juve_Deelman_2011} \citeS{Jackson_Muriki_2011} \citeS{Jackson_Ramakrishnan_2010} \citeS{Khamra_Kim_2010} \citeS{Li_Yang_2010} \citeS{Li_Humphrey_2010} \citeS{Lu_Jackson_2010} \citeS{Luckow_Jha_2010} \citeS{Lenk_Menzel_2011} \citeS{Liu_Zhai_2011} \citeS{Montella_Foster_2010} \citeS{Napper_Bientinesi_2009} \citeS{Ostermann_Iosup_2009} \citeS{Palankar_Iamnitchi_2008} \citeS{Redekopp_Simmhan_2011} \citeS{Rehr_Vila_2010} \citeS{Schatz_Koschnicke_2011} \citeS{Subramanian_Ma_2011} \citeS{Tran_Cinquini_2011} \citeS{Vockler_Juve_2011} \citeS{Vozmediano_Montero_2011} \citeS{Vecchiola_Pandey_2009} \citeS{Walker_2008} \citeS{Wall_Kudtarkar_2010} \citeS{Wilkening_Wilke_2009} \citeS{Zaspel_Griebel_2011}\\

Comparison between Computing Paradigms & \citeS{Carlyle_Harrell_2010} \citeS{Iosup_Ostermann_2011} \citeS{Kondo_Javadi_2009} \citeS{Zhai_Liu_2011}\\
\hline

\hline
\end{tabular}
\end{table}

According to these four evaluation purposes, the reviewed primary studies can be differentiated into four categories, as listed in Table \ref{tbl>6}. Note that one primary study may have more than one evaluation purposes, and we judge evaluation purposes of a study through its described application scenarios. For example, although the detailed evaluation contexts could be broad ranging from Cloud provider selection \citeS{Li_Yang_2010} to application feasibility verification \citeS{Vockler_Juve_2011}, we may generally recognize their purposes as \textit{Scientific Computing in the Cloud} if these studies investigated scientific applications in the Cloud. On the other hand, the studies like ``performance evaluation of popular Cloud IaaS providers" \citeS{Salah_Saba_2011} only have the motivation \textit{Cloud Resource Exploration} if they did not specify any application scenario. 

Apart from the evaluation work motivated by \textit{Cloud Resource Exploration}, we found that there are three times more attention paid to \textit{Scientific Computing in the Cloud} (40 studies) compared to \textit{Business Computing in the Cloud} (13 studies). In fact, the studies aiming at \textit{Comparison between Computing Paradigms} also intended to use Scientific Computing for their discussion and analysis \citeS{Carlyle_Harrell_2010,Kondo_Javadi_2009}. Given that Cloud Computing emerged as a business model \cite{Zhang_Cheng_2010}, public Cloud services are provided mainly to meet the technological and economic requirements from business enterprises, which does not match the characteristics of scientific computing workloads \citeS{He_Zhou_2010,Ostermann_Iosup_2009}. However, the study distribution over purposes (cf.~Table \ref{tbl>6}) suggests that the commercial Cloud Computing is still regarded as a potential and encouraging paradigm to deal with academic issues. We can find a set of reasons for this:
\begin{itemize*}
    \item	Since the relevant studies were all identified from academia (cf.~Section \ref{VII}), intuitively, Scientific Computing may seem more academic than Business Computing in the Cloud for researchers.
    \item	Although the public Cloud is deficient for Scientific Computing on the whole due to the relatively poor performance and significant variability \cite{Bientinesi_Iakymchuk_2010,Jackson_Ramakrishnan_2010,Ostermann_Iosup_2009}, smaller scale of computations can particularly benefit from the moderate computing capability of the Cloud \citeS{Carlyle_Harrell_2010,Hill_Humphrey_2009,Rehr_Vila_2010}.
    \item	The on-demand resource provisioning in the Cloud can satisfy some high-priority or time-sensitive requirements of scientific work when in-house resource capacity is insufficient \citeS{Carlyle_Harrell_2010,Hazelhurst_2008,Ostermann_Iosup_2009,Wilkening_Wilke_2009}.
    \item	It would be more cost effective to carry out temporary jobs on Cloud platforms to avoid the associated long-term overhead of powering and maintaining local computing systems \citeS{Carlyle_Harrell_2010,Ostermann_Iosup_2009}.
    \item	Through appropriate optimizations, the current commercial Cloud can be improved for Scientific Computing \citeS{Evangelinos_Hill_2008,Ostermann_Iosup_2009}.
    \item	Once commercial Cloud vendors pay more attention to Scientific Computing, they can make the current Cloud more academia-friendly by slightly changing their existing infrastructures \citeS{He_Zhou_2010}. Interestingly, the industry has acknowledged the academic requirements and started offering services for solving complex science/engineering problems \cite{Amazon_2011}.
\end{itemize*}

\subsection{RQ 2: What commercial Cloud services have been evaluated?}
\label{RQ2}
\begin{figure}[!t]
\centering
\includegraphics[width=8.5cm]{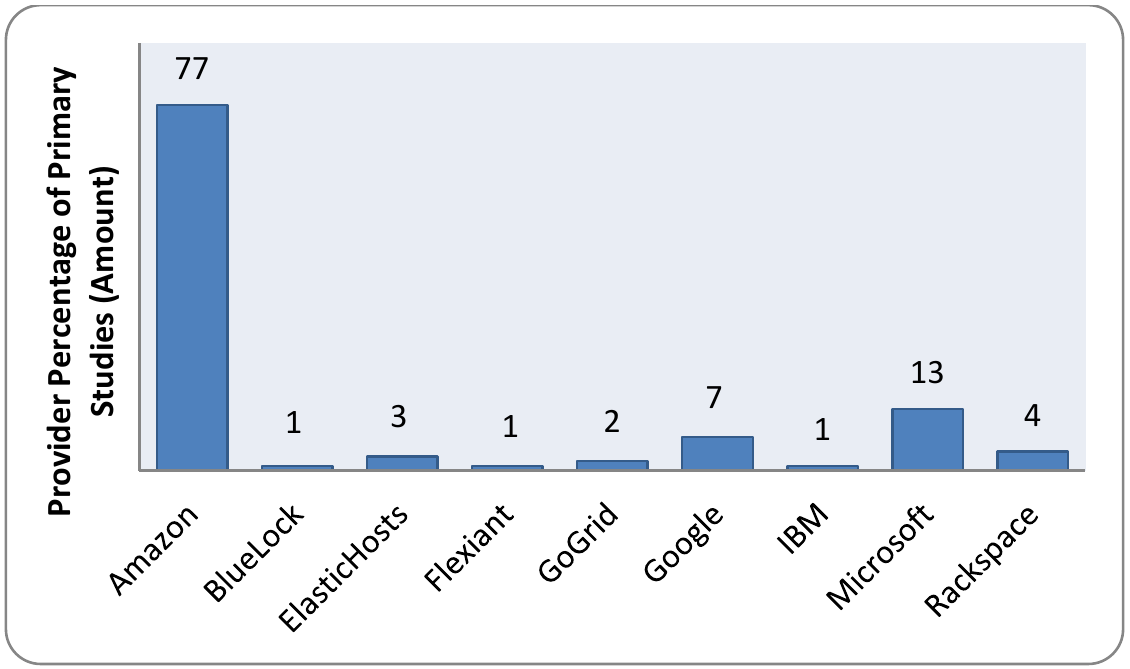}
\caption{\label{fig>vendors}Distribution of primary studies over Cloud providers.}
\end{figure}

Evaluations are based on services available from specific Cloud providers. Before discussing the individual Cloud services, we identify the service providers. Nine commercial Cloud providers have been identified in this SLR: Amazon, BlueLock, ElasticHosts, Flexiant, GoGrid, Google, IBM, Microsoft, and Rackspace. Mapping the 82 primary studies to these nine providers, as shown in Fig.\ \ref{fig>vendors}, we show that the commercial Cloud services attracting most evaluation efforts are provided by Amazon. Note that one primary study may cover more than one Cloud provider. This phenomenon is reasonable because Amazon has been treated as one of the top and key Cloud Computing providers in both industry and academia \cite{Buyya_Yeo_Venugopal_2009,Zhang_Cheng_2010}.

\begin{figure}[!t]
\centering
\includegraphics[width=8.5cm]{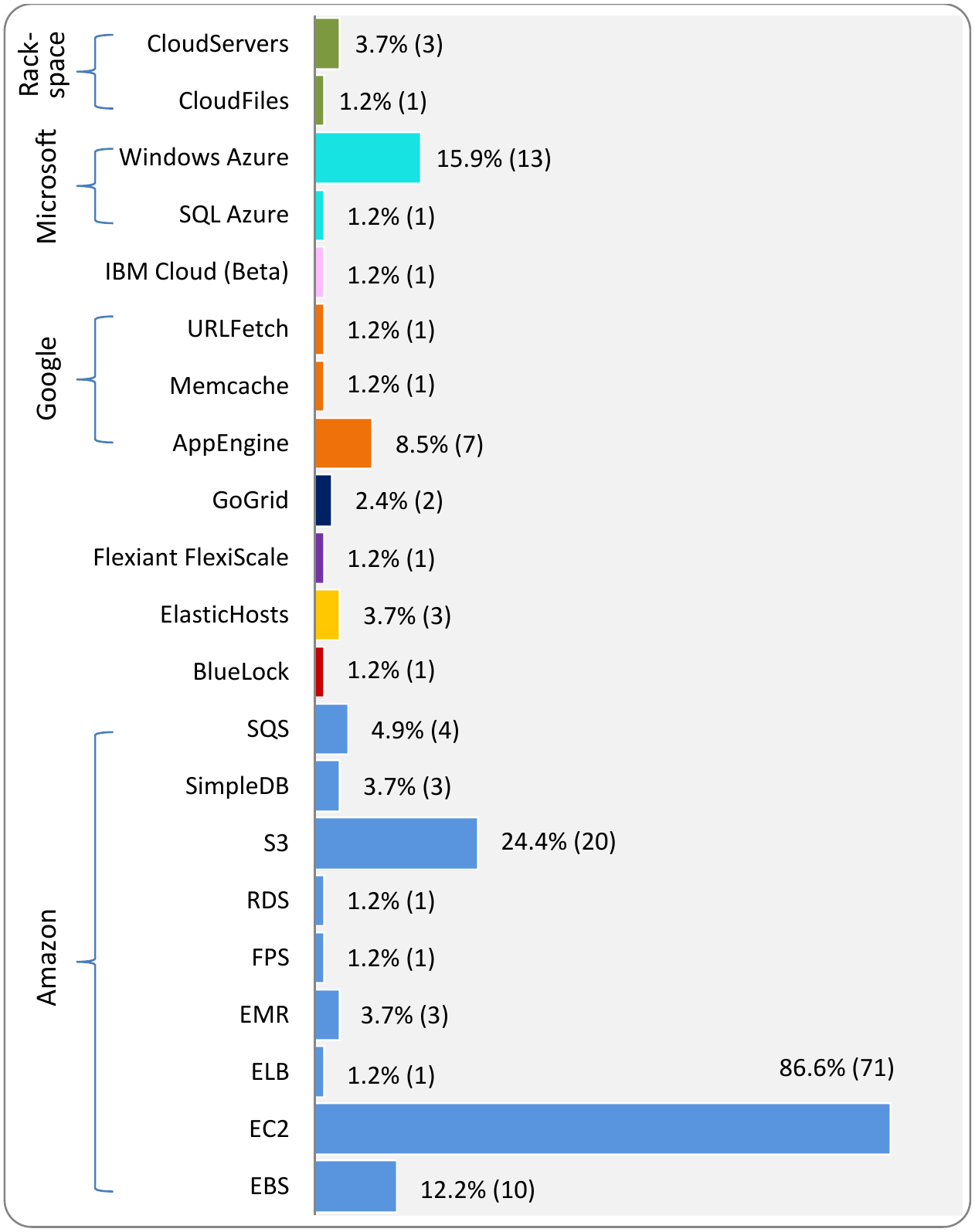}
\caption{\label{fig>Services}Distribution of primary studies over Cloud services.}
\end{figure}

With different public Cloud providers, we have explored the evaluated Cloud services in the reviewed publications, as listed in \ref{Appendix>II}. Note that the Cloud services are identified according to their commercial definitions instead of functional descriptions. For example, the work \citeS{Hill_Li_2010} explains Azure Storage Service and Azure Computing Service respectively, whereas we treated them as two different functional resources in the same Windows Azure service. The distribution of reviewed publications over detailed services is illustrated as shown in Fig.\ \ref{fig>Services}. Similarly, one primary study may perform evaluation of multiple commercial Cloud services. In particular, five services (namely Amazon EBS, EC2 and S3, Google AppEngine, and Microsoft Windows Azure) were the most frequently evaluated services compared with the others. Therefore, they can be viewed as the representative commercial Cloud services, at least in the context of Cloud services evaluation. Note that bias could be involved in the service identification in this work due to the pre-specified providers in the search string, as explained in Subsection \ref{VII>completeness_threat}.

Among these typical commercial Cloud services, Amazon EBS, EC2 and S3 belong to IaaS, Google AppEngine is PaaS, while Microsoft Windows Azure is recognized as a combination of IaaS and PaaS \citeS{Zhao_Liu_2010}. IaaS is the on-demand provisioning of infrastructural computing resources, and the most significant advantage is its flexibility \citeX{Binnig_Kossmann_2009}. PaaS refers to the delivery of a platform-level environment including operating system, software development frameworks, and readily available tools, which limits customers' control while taking complete responsibility of maintaining the environment on behalf of customers \citeX{Binnig_Kossmann_2009}. The study distribution over services (cf.~Fig.~\ref{fig>Services}) indicates that IaaS attracts more attention of evaluation work than PaaS. Such a finding is essentially consistent with the previous discussions when answering RQ1. The flexible IaaS may better fit into the diverse Scientific Computing. In fact, niche PaaS and SaaS are designed to provide additional benefits for their targeting applications, while IaaS is more immediately usable for particular and sophisticated applications \citeS{Juve_Deelman_2011}\cite{Harris_2012}. In other words, given the diversity of requirements in the Cloud market, IaaS and PaaS would serve different types of customers, and they cannot be replaced with each other. This finding can also be confirmed by a recent industry event: the traditional PaaS provider Google just offered a new IaaS -- Compute Engine \cite{Google_2012}. 

\subsection{RQ 3: What aspects and their properties of commercial Cloud services have been evaluated?}
\label{RQ3}
The aspects of commercial Cloud services can be initially investigated from general surveys and discussions about Cloud Computing. In brief, from the view of Berkeley \cite{Armbrust_Fox_2010}, Economics of Cloud Computing should be particularly emphasized in deciding whether to adopt Cloud or not. Therefore, we considered Economics as an aspect when evaluating commercial Cloud services. Meanwhile, although we do not agree with all the parameters identified for selecting Cloud Computing/Provider in \cite{Habib_Ries_2010}, we accepted Performance and Security as two significant aspects of a commercial Cloud service. Such an initial investigation of service aspects has been verified by this SLR. Only Performance, Economics, and Security and their properties have been evaluated in the primary studies. 

\begin{figure}[!t]
\centering
\includegraphics{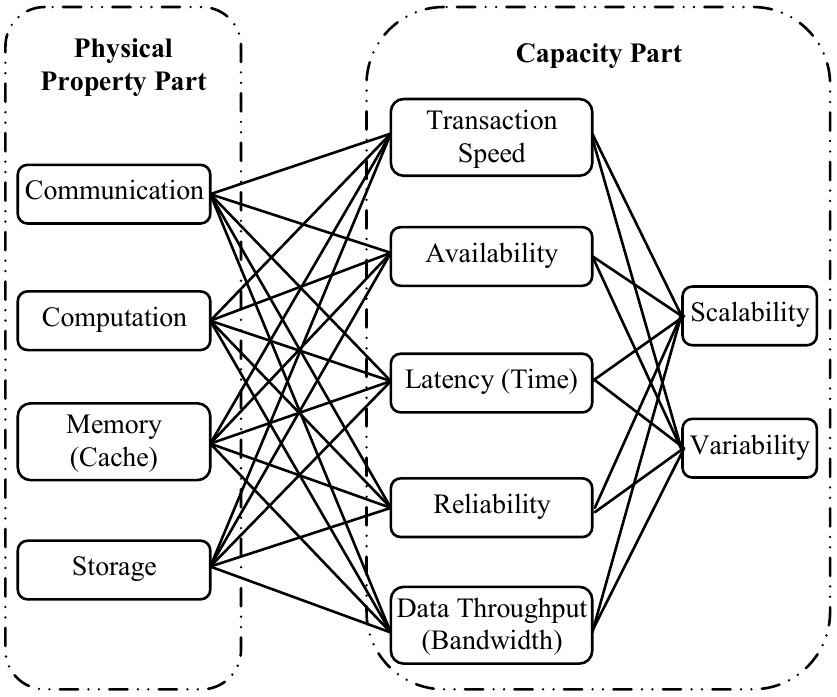}
\caption{\label{fig>PerformanceProperties}The properties of the Performance aspect (from \cite{Li_OBrien_2012a}).}
\end{figure}

The detailed properties and the corresponding distribution of primary studies are listed in Table \ref{tbl>7}. Note that a primary study usually covers multiple Cloud service aspects and/or properties. In particular, we only take into account the physical properties for the Performance aspect in this paper. The capacities of different physical properties and their sophisticated correlations (cf.~Fig.~\ref{fig>PerformanceProperties}) have been specified in our previous work \cite{Li_OBrien_2012a}. 

\begin{table}[!t]\footnotesize
\renewcommand{\arraystretch}{1.3}
\centering
\caption{\label{tbl>7}Distribution of studies over Cloud service aspects/properties.}
\begin{tabular}{l p{2.8cm} l l}
\hline

\hline
Aspect & Property & \#Papers & Percentage\\
\hline
\multirow{6}{1.7cm}{Performance} & Communication & 24 & 29.27\%\\
& Computation & 20 & 24.39\%\\
& Memory(Cache) & 12 & 14.63\%\\
& Storage & 28 & 34.15\%\\
& Overall Performance & 48 & 58.54\%\\
\cline{2-4}
& Total & 78 & 95.12\%\\
\hline
\multirow{3}{1.7cm}{Economics} & Cost & 35 & 42.68\%\\
& Elasticity & 9 & 10.98\%\\
\cline{2-4}
& Total & 40 & 48.78\%\\
\hline
\multirow{5}{1.7cm}{Security} & Authentication & 1 & 1.22\%\\
& Data Security & 4 & 4.88\%\\
& Infrastructural Security & 1 & 1.22\%\\
& Overall Security & 1 & 1.22\%\\
\cline{2-4}
& Total & 6 & 7.32\%\\
\hline

\hline
\end{tabular}
\end{table}

Overall, we find that the existing evaluation work overwhelmingly focused on the performance features of commercial Cloud services. Many other theoretical concerns about commercial Cloud Computing, Security in particular, were not well evaluated yet in practice. Given the study distribution over service aspects/properties (cf.~Table \ref{tbl>7}), several research gaps can be revealed or confirmed:
\begin{itemize*}
    \item	Since memory/cache could closely work with the computation and storage resources in computing jobs, it is hard to exactly distinguish the effect to performance brought by memory/cache, which may be the main reason why few dedicated Cloud memory/cache evaluation studies were found from the literature. In addition to the memory performance, the memory hierarchy could be another interesting issue to be evaluated \citeS{Ostermann_Iosup_2009}.
    \item	Although one major benefit claimed for Cloud Computing is elasticity, it seems difficult for people to know how elastic a Cloud platform is. In fact, evaluating elasticity of a Cloud service is not trivial \cite{Kossmann_Kraska_2010_j}, and there is little explicit measurement to quantify the amount of elasticity in a Cloud platform \cite{Islam_Lee_2012}.
    \item	The security of commercial Cloud services has many dimensions and issues people should be concerned with \cite{Armbrust_Fox_2010,Zhang_Cheng_2010}. However, not many security evaluations were reflected in the identified primary studies. Similar to the above discussion about elasticity evaluation, the main reason may be that the security is also hard to quantify \cite{Brooks_2012}. Therefore, we conclude that the Elasticity and Security evaluation of commercial Cloud services could be a long-term research challenge.
\end{itemize*}

\begin{table}[!t]\footnotesize
\renewcommand{\arraystretch}{1.3}
\centering
\caption{\label{tbl>8}Distribution of metrics over Cloud service aspects/properties (based on \cite{Li_OBrien_2012b} and updated).}
\begin{tabular}{l l l}
\hline

\hline
Aspect & Property & \#Metrics\\
\hline
\multirow{5}{*}{Performance} & Communication & 9\\
& Computation & 7\\
& Memory (Cache) & 7\\
& Storage & 11\\
& Overall Performance & 18\\
\hline
\multirow{2}{*}{Economics} & Cost & 18\\
& Elasticity & 4\\
\hline
\multirow{4}{*}{Security} & Authentication & 1\\
& Data Security & 3\\
& Infrastructural Security & 1\\
& Overall Security & 1\\
\hline

\hline
\end{tabular}
\end{table}

\begin{table*}[!t]\footnotesize
\renewcommand{\arraystretch}{1.3}
\centering
\caption{\label{tbl>benchmark}The traditional benchmarks used in Cloud services evaluation.}
\begin{tabular}{l l c | c c c c c}
\hline

\hline
\multirow{2}{*}{Benchmark}&\multirow{2}{*}{Type} &\multirow{2}{*}{Applicability} & \multicolumn{5}{c}{Evaluated Cloud Service Property (with one study as a sample)}\\
\cline{4-8}
 & &  & Communication & Computation & Memory/Cache & Storage & Overall Performance\\
\hline
An Astronomy workflow & Application & 1 & & & & &   \citeS{Vockler_Juve_2011}\\
Application/Workflow Suite & Application & 3 &  \citeS{Jackson_Ramakrishnan_2010}& \citeS{Dejun_Pierre_2010}& & &   \citeS{Jackson_Ramakrishnan_2010}\\
B+\_Tree indexing system & Application & 1 & & & \citeS{Chiu_Hall_2011} & \citeS{Chiu_Hall_2011} &\\
Badabing Tool & Micro & 1 &  \citeS{Wang_Ng_2010}& & & & \\
Betweenness Centrality & Application & 1 & & & & &    \citeS{Redekopp_Simmhan_2011} \\
BitTorrent & Application & 1 & & &  &   \citeS{Palankar_Iamnitchi_2008} & \\
BLAST/BLAST+ & Application & 6 & & &  & &    \citeS{Lu_Jackson_2010}\\
Bonnie/Bonnie++ & Micro& 4 &  & &    \citeS{Ostermann_Iosup_2009}    &   \citeS{Ostermann_Iosup_2009} & \\
Broadband & Application & 3 & & &\citeS{Juve_Deelman_2011} & &    \citeS{Juve_Deelman_2009} \\
CacheBench & Micro & 2 & & &    \citeS{Ostermann_Iosup_2009}    & & \\
CAP3 & Application & 1 & & & & &    \citeS{Gunarathne_Wu_2011} \\
Classify gene data & Application & 1 & & & & & \citeS{Vecchiola_Pandey_2009} \\
Compiling Linux Kernel & Application & 1 & &    \citeS{Baun_Kunze_2009} & & & \\
CSFV & Application & 1& & & & &    \citeS{He_Zhou_2010} \\
Dhrystone & Synthetic & 1 & & \cite{Phillips_Engen_2011} & & & \\
EnKF-based matching & Application & 1 & & & & &    \citeS{Khamra_Kim_2010} \\
Epigenome & Application &3 & & \citeS{Juve_Deelman_2011} & & &    \citeS{Juve_Deelman_2009} \\
FEFF84 MPI & Application &1 & & & & &    \citeS{Rehr_Vila_2010} \\
Fibonacci & Micro &1 & &    \citeS{Iosup_Yigitbasi_2011} & & & \\
FIO & Micro &1 & & & &\citeS{Salah_Saba_2011} &  \\
fMRI brain imaging & Application & 1 & & & & & \citeS{Vecchiola_Pandey_2009} \\
GASOLINE & Application & 1& &  & & &     \citeS{Rehr_Vila_2010}\\
Grapes & Application& 1 & & &   &  & \citeS{Zhai_Liu_2011}\\
GTM & Application & 1 & & & & &    \citeS{Gunarathne_Wu_2011} \\
Hadoop App & Application &2 & & & & &    \citeS{Dalman_Doernemann_2010} \\
hdparm tool & Synthetic & 1  & & &   & \citeS{Zhai_Liu_2011} & \\
HPCC: b\_eff & Micro & 3 &   \citeS{Ostermann_Iosup_2009}& & & & \\
HPCC: DGEMM & Micro & 5 & &    \citeS{Jackson_Ramakrishnan_2010} &    \citeS{Bientinesi_Iakymchuk_2010}  & & \\
HPCC: FFTE & Synthetic & 1& &    \citeS{Jackson_Ramakrishnan_2010} & & & \\
HPCC: HPL & Synthetic & 8 &  &   \citeS{Ostermann_Iosup_2009} &   \citeS{Bientinesi_Iakymchuk_2010} & &   \citeS{Akioka_Muraoka_2010}\\
HPCC: PTRANS & Synthetic & 1 &  \citeS{Jackson_Ramakrishnan_2010}& & & & \\
HPCC: RandomAccess & Synthetic & 3 & & &    \citeS{Jackson_Ramakrishnan_2010}    & & \\
HPCC: STREAM & Micro & 6 & & &    \citeS{Ostermann_Iosup_2009}    & & \\
iperf & Micro & 4 &   \citeS{Li_Yang_2010}& & & & \\
Intel MPI Bench & Micro & 3&   \citeS{Hill_Humphrey_2009}& & & & \\
IOR & Synthetic & 4 & & &  \citeS{Ghoshal_Canon_2011}  &   \citeS{Evangelinos_Hill_2008}  & \\
Isabel & Application & 1 & \citeS{Cervino_Rodriguez_2011} & & & & \\
KMeans Clustering & Application & 1 & & & & &    \citeS{Bicer_Chiu_2011} \\
Land Elevation Change & Application  &1 & & &    \citeS{Chiu_Agrawal_2010}  & & \\
Latency Sensitive Website & Application &1 &   \citeS{Li_Yang_2010}& & & & \\
Livermore Loops & Synthetic & 1 & & \cite{Phillips_Engen_2011} & & & \\
LMbench  & Micro & 4 & &    \citeS{Jayasinghe_Malkowski_2011} & & & \citeS{Iosup_Ostermann_2011}\\
Lublin99 & Synthetic &1 & & & & &    \citeS{Assuncao_Costanzo_2010} \\
MapReduce App & Application & 1& & & & &    \citeS{Schad_Dittrich_2010} \\

MG-RAST + BLAST & Application & 1 & & & & &    \citeS{Wilkening_Wilke_2009} \\
Minion Constraint solver & Application & 1 & & & & &    \citeS{Gent_Kotthoff_2011} \\
mpptest & Micro & 1&   \citeS{He_Zhou_2010}& & & & \\
MODIS Processing & Application & 2 & & & & &    \citeS{Li_Humphrey_2010} \\
Montage & Application &4 & & & & \citeS{Juve_Deelman_2011} &    \citeS{Juve_Deelman_2009} \\
NaSt3DGPF & Application & 1 & & & & &    \citeS{Zaspel_Griebel_2011} \\

\multicolumn{8}{r}{Continued on next page} \\

\end{tabular}
\end{table*}

\begin{table*}[!t]\footnotesize
\renewcommand{\arraystretch}{1.3}
\centering
\begin{tabular}{l l c | c c c c c}
\multicolumn{8}{l}{Table \ref{tbl>benchmark} (continued from previous page)} \\
\hline

\hline
\multirow{2}{*}{Benchmark}&\multirow{2}{*}{Type} &\multirow{2}{*}{Applicability} & \multicolumn{5}{c}{Evaluated Cloud Service Property (with one study as a sample)}\\
\cline{4-8}
 & &  & Communication & Computation & Memory/Cache & Storage & Overall Performance\\
\hline

NetPIPE & Micro &1 & \citeS{Jayasinghe_Malkowski_2011} & & & & \\
NPB: BT & Synthetic & 2 & & &   &   \citeS{Akioka_Muraoka_2010}   & \\
NPB: BT-IO & Synthetic & 2 & & &   &   \citeS{Evangelinos_Hill_2008}   & \\
NPB: EP & Micro &1 & &    \citeS{Akioka_Muraoka_2010} & & & \\
NPB: GridNPB: ED & Synthetic & 1 & & & & &    \citeS{Vozmediano_Montero_2011} \\
NPB: original & Synth+Micro & 4 & \citeS{Zhai_Liu_2011} &    \citeS{Carlyle_Harrell_2010} & & &   \citeS{Akioka_Muraoka_2010}\\
NPB-OMP & Synthetic & 2& & & & &    \citeS{Walker_2008} \\
NPB-MPI & Synthetic & 2&\citeS{He_Zhou_2010} & & & &    \citeS{Walker_2008} \\
NPB-MZ & Synthetic & 1& & & & &    \citeS{He_Zhou_2010} \\
OMB-3.1 with MPI & Micro &1 &   \citeS{Evangelinos_Hill_2008}& & & & \\
Operate/Transfer Data & Micro & 19 &\citeS{Baun_Kunze_2009} & &    &   \citeS{Li_Yang_2010} & \\
PageRank & Application & 1 & & & & &    \citeS{Bicer_Chiu_2011} \\
Passmark CPU Mark & Micro & 1 & & \citeS{Lenk_Menzel_2011} & & &\\
PCA & Application & 1 & & & & &    \citeS{Bicer_Chiu_2011} \\
Phoronix Test Suite & Application  & 1 & & & & &    \citeS{Lenk_Menzel_2011} \\
ping & Micro & 5 &   \citeS{Li_Yang_2010}& & & & \\
POP  & Application & 2 & & &   &  \citeS{Liu_Zhai_2011}  & \citeS{Zhai_Liu_2011}\\
PostMark & Synthetic &1 & & &    &   \citeS{Wang_Varman_2011}  & \\
ROIPAC workflow & Application & 1 & & & & &    \citeS{Tran_Cinquini_2011} \\
RUBBoS+MySQL Cluster & Application & 1 & & & & &    \citeS{Jayasinghe_Malkowski_2011} \\
SAGA BigJob System & Application & 1& & & & &    \citeS{Luckow_Jha_2010} \\
Seismic Source Inversion & Application & 1 & & & & &    \citeS{Subramanian_Ma_2011} \\
Simplex & Micro & 1 & & \citeS{Salah_Saba_2011} & & &\\
SNfactory & Application & 1 & \citeS{Jackson_Muriki_2011} & \citeS{Jackson_Muriki_2011} && \citeS{Jackson_Muriki_2011} & \citeS{Jackson_Muriki_2011}\\
Social Website & Application & 1 & & & & &    \citeS{Ramasahayam_Deters_2011} \\
SPECjvm 2008 & Synthetic  & 1& & & & &    \citeS{Li_Yang_2010} \\
SPECweb & Synthetic & 2 &  \citeS{Liu_Wee_2009}&   \citeS{Liu_Wee_2009} & & & \citeS{Chen_Bai_2011}\\
Sysbench on MySQL & Application  & 1& & &    & &    \citeS{Sobel_Subramanyam_2008} \\
Timed Benchmark & Synthetic & 1 & & &   &   \citeS{Ghoshal_Canon_2011}   & \\
TORCH Benchmark Suite & Synthetic & 1 & & &   &    & \citeS{Phillips_Engen_2011} \\
TPC-E  & Synthetic & 1& & & & &    \citeS{Hill_Li_2010} \\
TPC-W & Synthetic & 4 & & &    &   \citeS{Li_Yang_2010}  &   \citeS{Kossmann_Kraska_2010}\\
Ubench & Micro & 1& &    \citeS{Schad_Dittrich_2010} &    \citeS{Schad_Dittrich_2010} & & \\
WCD & Application &1 & & & & &    \citeS{Hazelhurst_2008} \\
Whetstone & Synthetic  & 1& &    \citeS{Kondo_Javadi_2009} & & & \\
WSTest & Synthetic  &1 & & & & &    \citeS{Stantchev_2009} \\
\hline

\hline
\end{tabular}
\end{table*}

\subsection{RQ 4: What metrics have been used for evaluation of commercial Cloud services?}
\label{RQ4}
Benefiting from the above investigation of aspects and their properties of commercial Cloud services, we can conveniently identify and organize their corresponding evaluation metrics. In fact, more than 500 metrics including duplications have been isolated from the experiments described in the primary studies. After removing the duplications, we categorized and arranged the metrics naturally following the aforementioned Cloud service aspects/properties. Note that we judged duplicate metrics according to their usage contexts instead of names. Some metrics with different names could be essentially duplicate ones, while some metrics with identical name should be distinguished if they are used for different evaluation objectives. For example, the metric \textit{Upload/Download Data Throughput} has been used for evaluating both Communication \citeS{Hazelhurst_2008} and Storage \citeS{Palankar_Iamnitchi_2008}, and therefore it was arranged under both Cloud service properties.

Due to the limit of space, we do not elaborate all the identified metrics in this paper. In fact, we have summarized the existing evaluation metrics into a catalogue to facilitate the future practice and research in the area of Cloud services evaluation \cite{Li_OBrien_2012b}. Here we only give a quick impression of their usage by displaying the distribution of those metrics, as shown in Table \ref{tbl>8}.

Given the distribution together with the catalogue of Cloud services evaluation metrics, we summarize several findings below:
\begin{itemize*}
    \item	The existing evaluation work has used a large number of metrics to measure various performance features as well as the cost of commercial Cloud services. This confirms the current fashion of cost evaluation: based on performance evaluation, evaluators analyze and estimate the real expense of using Cloud services \citeS{Lenk_Menzel_2011,Zhai_Liu_2011}. We may name this type of evaluated cost as resource cost. In fact, the cost of Cloud Computing may cover a wide range of theoretical concerns, such as migration cost, operation cost, etc. \cite{Armbrust_Fox_2010}. However, those costs depend on specific systems, technologies, human activities, and even environmental factors. Performing generic cost evaluation could then be a tremendous challenge. A promising solution to this challenge is to replace the cost with other steady factors for evaluation. For example, we may estimate the size of Cloud migration projects instead of directly evaluating the migration cost \cite{Tran_Lee_2011}.
    \item	There is still a lack of effective metrics for evaluating Cloud elasticity. As mentioned previously, it is not easy to explicitly quantify the amount of elasticity of a Cloud service. To address this research gap, as far as we know, the most recent effort is a sophisticated Penalty Model that measures the imperfections in elasticity of Cloud services for a given workload in monetary units \cite{Islam_Lee_2012}.
    \item	It seems that there is no suitable metric yet to evaluate security features of Cloud services, which also confirms the previous findings in Section \ref{RQ3}. Since security is hard to quantify \cite{Brooks_2012}, current security evaluation has been realized mainly by qualitative discussions. A relatively specific suggestion for security evaluation of Cloud services is given in \citeS{Palankar_Iamnitchi_2008}: the security assessment can start with an evaluation of the involved risks. As such, we can use a pre-identified risk list to discuss the security strategies supplied by Cloud services.
\end{itemize*}

\subsection{RQ 5: What benchmarks have been used for evaluation of commercial Cloud services?}
This SLR has identified around 90 different benchmarks in the selected studies of Cloud services evaluation. As discussed in the related work (cf.~Section \ref{II}), there are several emerging and dedicated Cloud benchmarks, such as YCSB \citeX{Cooper_Silberstein_2010}, CloudStone \citeS{Sobel_Subramanyam_2008}, and CloudSuite \cite{Ferdman_Adileh_2012}. Traditional benchmarks have still been overwhelmingly used in the existing practices of Cloud services evaluation, as summarized in Table \ref{tbl>benchmark}. Note that, in Table \ref{tbl>benchmark}, each benchmark together with a corresponding evaluated service property cites only one relevant study as an instance. In particular, the evaluated Economics and Security properties are not reflected in this table. First, the existing cost evaluation studies were generally based on the corresponding performance evaluation \citeS{Lenk_Menzel_2011,Zhai_Liu_2011}. Second, the selected studies did not specify any distinct benchmark for evaluating elasticity and security. Through Table \ref{tbl>benchmark} we show that, although the traditional benchmarks were recognized as being insufficient for evaluating commercial Cloud services \citeX{Binnig_Kossmann_2009}, traditional benchmarks can still satisfy at least partial requirements of Cloud services evaluation.

Moreover, one benchmark may be employed in multiple evaluation practices. The numerous evaluators' experiences can then be used to indicate the applicability of a particular benchmark. Here we define a benchmark's ``Applicability" as the number of the related studies. Through the applicability of different traditional benchmarks (cf.~Table \ref{tbl>benchmark}), we list the popular benchmarks as recommendations for Cloud services evaluation, as shown in Table \ref{tbl>PopularBenchmark}. 
\begin{table}[!t]\footnotesize
\renewcommand{\arraystretch}{1.3}
\centering
\caption{\label{tbl>PopularBenchmark}Popular traditional benchmarks for evaluating different Cloud service properties.}
\begin{tabular}{l p{5cm}}
\hline

\hline
Cloud Service Property & Popular Traditional Benchmarks\\
\hline
Communication & iperf, ping, Operate/Transfer Data\\
Computation & HPCC: DGEMM, HPCC: HPL, LMBench\\
Memory/Cache & HPCC: STREAM\\
Storage & Bonnie/Bonnie++, IOR, NPB: BT/BT-IO, Operate/Transfer Data\\
Overall Performance & BLAST, HPCC: HPL, Montage, NPB suite, TPC-W\\
\hline

\hline
\end{tabular}
\end{table}

In addition, following the evolution of benchmarking in the computing area \cite{Lewis_Crews_1985}, we summarized three types of benchmarks used for evaluating commercial Cloud services: Application Benchmark, Synthetic Benchmark, and Micro-Benchmark.

\begin{itemize*}
    \item	Application Benchmark refers to the real-world software systems that are deployed to the Cloud and used as potentially true measures of commercial Cloud services.
    \item	Synthetic Benchmark is not a real application, but a well-designed program using representative operations and workload to simulate a typical set of applications.
    \item	Micro-Benchmark is a relatively simple program that attempts to measure a specific component or a basic feature of Cloud services.
\end{itemize*}

To give a quick impression of what types of benchmarks were adopted in the current Cloud services evaluation work, we list the distribution of primary studies over employed benchmark types, as shown in Table \ref{tbl>9}.
\begin{table}[!t]\footnotesize
\renewcommand{\arraystretch}{1.3}
\centering
\caption{\label{tbl>9}Distribution of studies over benchmark types.}
\begin{tabular}{l l l}
\hline

\hline
Benchmark Type & \#Papers & Percentage\\
\hline
Application Only & 27 & 32.93\%\\
Synthetic Only & 11 & 13.41\%\\
Micro Only & 17 & 20.73\%\\
Application + Synthetic & 3 & 3.66\%\\
Application + Micro & 12 & 14.63\%\\
Synthetic + Micro & 6 & 7.32\%\\
All Three & 6 & 7.32\%\\
\hline
Total & 82 & 100\%\\
\hline

\hline
\end{tabular}
\end{table}

It can be seen that more than half of the primary studies adopted only one particular type of benchmark to evaluate commercial Cloud services. Given that different types of benchmarks reveal different service natures, it is impossible to use one benchmark to fit all when performing Cloud services evaluation. Thus, a recommendation from this SLR is to employ a suite of mixed types of benchmarks to evaluate Cloud services in the future.

\subsection{RQ 6: What experimental setup scenes have been adopted for evaluating commercial Cloud services?}
\label{RQ6}
\begin{figure*}[!t]
\centering
\includegraphics{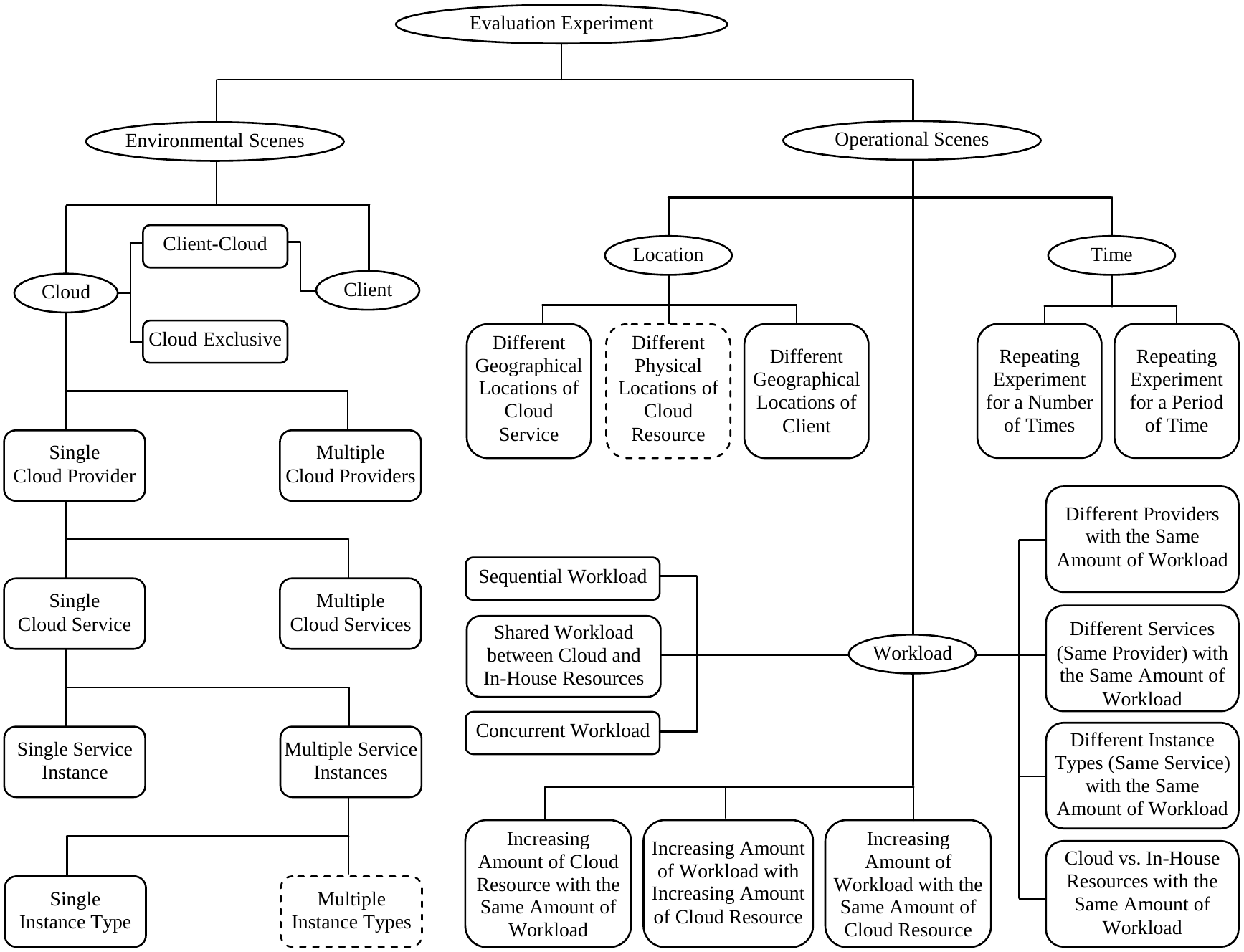}
\caption{\label{fig>PicScene}Experimental setup scene tree of performance evaluation of commercial Cloud services (from \cite{Li_OBrien_2012a}).}
\end{figure*}

As mentioned in Section \ref{III>RQs}, we used ``setup scene" to indicate an atomic unit for constructing complete Cloud services evaluation experiments. Through extracting different data from a primary study for respectively answering the data extraction questions (12) and (13) (cf.~Subsection \ref{III>schema}), we can distinguish between environmental setup scenes and operational setup scenes. The environmental setup scenes indicate static descriptions used to specify required experimental resources, while the operational setup scenes indicate dynamic operations that usually imply repeating an individual experiment job under different circumstances. For the convenience of analysis, the operational setup scenes were further divided into three groups with respect to experimental Time, Location, and Workload. In detail, ten environmental setup scenes and 15 operational setup scenes have been identified, which can be organized as an experimental setup scene tree, as shown in Fig.~\ref{fig>PicScene}.

We have developed a taxonomy to clarify and structure these 25 experimental setup scenes in a separate piece of work \cite{Li_OBrien_2012a}. In particular, the rounded rectangle with dashed line (Fig.~\ref{fig>PicScene}) represents the setup scenes that are either uncontrollable (\textit{Different Physical Locations of Cloud Resource}) or unemployed yet (\textit{Multiple Instance Types}). The physical location of a particular Cloud resource indicates its un-virtualized environment. The un-virtualized difference then refers not only to the difference in underlying hardware like different model of real CPU, but also to the difference between VMs sharing or not sharing underlying hardware. As for the setup scene \textit{Multiple Instance Types}, although it is possible to assign different functional roles to different types of VM instances to finish a single experiment job, we have not found such jobs in the reviewed literature.

Overall, by using the experimental setup scene tree, we can easily locate or enumerate individual environmental and operational setup scenes for Cloud services evaluation studies. As such, the answer to this research question may be employed essentially to facilitate drawing experimental lessons from the existing evaluation reports, and to facilitate the evaluation-related communication among the Cloud Computing community.

\section{Experiences of applying the SLR method}
\label{VI}
This SLR was prepared by a review team and two consultants, implemented primarily by a PhD student under supervision, and discussed and finalized by the whole team. According to our practice of conducting this study, we summarized some experiences to which or against which researchers can refer or debate in future SLR implementations.

First of all, a question-oriented SLR is apparently more efficient than an ad hoc review. For a new comer in a particular research area, it is difficult to measure his/her study progress if he/she is doing an ad hoc literature review. On the contrary, benefiting from the SLR, the progress becomes traceable by following a standardized procedure \cite{Kitchenham_Charters_2007}.

However, it should be noticed that traditional ad hoc reviews cannot be completely replaced with SLRs. Although supervisors can help introduce the background and/or motivation in advance, it is crucial for the student to comprehend enough relevant domain knowledge before starting an SLR. In terms of our experience with this SLR, an ad hoc review still showed its value in obtaining domain knowledge in a short period, which confirms that it is necessary to \textquotedblleft thoroughly understand the nature and scale of the task at hand before undertaking a SLR\textquotedblright { }\cite{Major_Kyriacou_2011}. When an SLR is supposed to be implemented by PhD students in an unfamiliar area, we should also estimate and consider the additional time on students' traditional review.

Moreover, our study also confirmed that a pilot review is vital for an SLR \cite{Barbar_Zhang_2009}. The pilot review of an SLR can be viewed as a bridge between the SLR and the corresponding ad hoc review. On one hand, the pilot review can reinforce or revise the reviewers' comprehension of domain-specific knowledge. On the other hand, the pilot review can help refine research questions, improve search strategy, and verify data extraction schema by trying to answer research questions. Therefore, we suggest that a pilot review can be done together with constructing the SLR protocol.

Additionally, for some research topics, the employment of an SLR is worthy of regular use to keep the relevant data or knowledge current to support those topics. According to Zhang and Babar's survey \cite{Zhang_Babar_2011}, most of existing SLRs in software engineering area seem one-off studies, such as to outline state-of-the-art or to get knowledge within a particular research region. Whereas, for this study, we plan to use the collected data to fill an experience base to support a Cloud services evaluation methodology. Considering the knowledge in an expert system should be updated regularly, it is necessary to always keep the corresponding experience base up to date. In this case, therefore, we will continually collect relevant primary studies, and periodically update this SLR work.

Overall, in this study, the SLR method has been verified suitable and helpful for a first-year PhD student to accumulate knowledge and identify his research opportunities.

\section{Threats to validity}
\label{VII}
Although we tried to conduct this SLR study as rigorously as possible, it may have still suffered from several validity threats, as listed below. The future work should take into account these limitations when interpreting or directly using the findings or conclusions in this report.

\subsection{Conceptual model of Cloud services evaluation}
\label{VII>model_threat}
The construction of this SLR and the following investigation into Cloud services evaluation were based on the proposed conceptual model (cf.~Section \ref{II}). Therefore, any inaccuracy in the conceptual model of Cloud services evaluation may bring flaws in this study. As previously mentioned, we built this conceptual model by adapting a systematic performance evaluation approach \cite{Jain_1991}. In particular, we deliberately ignored two steps in the general process of evaluation implementation, namely evaluation factor identification and experimental result analysis. The reason for ignoring the former, we found that it was hard to directly extract experimental factors from the primary studies. To the best of our knowledge, although the existing evaluation experiments essentially involved factors, none of the current Cloud evaluation studies specified ``experimental factors" \cite{Montgomery_2009} in advance to design evaluation experiments and analyze the experimental results. In fact, we finally investigated potential factors through a secondary analysis of the answer to RQ6 in this SLR \cite{Li_OBrien_2012c}. The reason for ignoring the latter, as mentioned in the Introduction, we conducted this SLR study to investigate the procedures and experiences of Cloud services evaluation rather than the evaluation results. Overall, although we are not aware of any bias introduced by this conceptual model, other researchers with different interest may have different opinions about the intentionally ignored information.

\subsection{Research scope}
\label{VII>scope_threat}
The practices of Cloud services evaluation are reported in various sources, such as academic publications, technical websites, blogs, etc. In particular, the academic publications are normally formal reports after rigorous peer reviewing. Considering the generally specific and precise documentation of evaluation implementations in formal publications \cite{Sarmad_Ali_2010}, we limited this SLR to academic studies only. There is no doubt that informal descriptions of Cloud services evaluation in blogs and technical websites can also provide highly relevant information. However, on the one hand, it is impossible to explore and collect useful data from different study sources all at once. On the other hand, the published evaluation studies can be viewed as typical representatives of the existing ad hoc evaluation practices. By using the SLR method to exhaustively investigate the academic studies, we are still able to rationally show the representative state-of-the-practice of the evaluation of commercial Cloud services. In fact, we proposed to use the result of this SLR to construct a knowledge base first. The knowledge base can be gradually extended and enriched by including the other informal empirical studies of Cloud services evaluation.

\subsection{Completeness}
\label{VII>completeness_threat}
Given the increasing number of studies in this area, we note that we cannot guarantee to have captured all the relevant studies. The possible reasons could be various ranging from the search engines to the search string. Firstly, we did not look into every possible search resource. To balance between the estimated workload and coverage, five electronic libraries were selected based on the existing SLR experiences (cf.~Section \ref{III>resources}). In fact, the statistics suggests that these five literature search engines may give a broad enough coverage of relevant studies \cite{Zhang_Babar_Tell_2011}. Secondly, we unfolded automated search through titles, keywords and abstracts instead of full texts. On one hand, using a full text search usually leads to an explosion of search result. On the other hand, the search precision would be reduced quite dramatically by scanning full texts \cite{Dieste_Griman_2009}. Thirdly, due to the known limitations of the search engines \cite{Brereton_Kitchenham_2007}, we also noticed and confirmed that the automated search missed important studies. To alleviate this issue, we supplemented a manual search by snowballing the references of the initially selected papers (cf.~Section \ref{III>process}). Fourthly, it is possible that we may have not found the papers using irregular terms to describe Cloud services evaluation. In addition to carefully proposing the search string (cf.~Section \ref{III>string}), similarly, we also resorted to the reference snowballing to further identify the possibly missed publications. Finally, we specified ten Cloud providers in the search string, which may result in bias when identifying the most common services and providers to answer RQ2. However, we had to adopt those search terms as a tradeoff for improving the search string's sensitivity of the ``commercial Cloud service"-related evaluation studies. Since the top ten Cloud providers were summarized by the third party from the industrial perspective, they can be viewed as weighted popular providers for this study. In fact, other Cloud providers were still able to be identified, such as BlueLock, EasticHosts, and Flexiant (cf.~Section \ref{RQ2}).

\subsection{Reviewers reliability}
As mentioned in Section \ref{III>role}, the detailed review work was implemented mainly by a PhD student to gain understanding of his research topic. Since the student is a new comer in the Cloud Computing domain, his misunderstanding of Cloud services evaluation may incur biased review process and results. To help ensure that the conduction of this SLR was as unbiased as possible, we adopted a supervisory strategy including three points: first, before planning this SLR, the supervisory panel instructed the PhD student to perform an ad hoc review of background knowledge covering Cloud Computing in general and Cloud services evaluation in particular; second, during planning this SLR, the expert panel was involved in helping develop a review protocol prior to conducting the review; third, every step of the conduction of this SLR was under close supervision including regular meetings, and all the unsure issues were further discussed with the expert panel. As such, we have tried our best to reduce the possible bias of the review conduction. However, when it comes to the data analysis, there might still be the possibility of incomplete findings or conclusions due to our personal interest and opinions.

\subsection{Data extraction}
During the process of data extraction from the reviewed studies, we found that not many papers specified sufficient details about the evaluation background, environment, and procedure, which could be partially reflected by the quality assessment. As a result, sometimes we had to infer certain information through some unclear clues, particularly when we tried to find the purpose or the time of particular evaluation experiments. Therefore, there may be some inaccuracies in the inferred data. However, this point can be considered as a limitation of the current primary studies instead of this SLR. Since the empirical research in Cloud services evaluation falls in the experimental computer science \cite{Feitelson_2007}, we suggest that researchers may employ structural abstract \cite{Budgen_Kitchenham_2008} and/or guidelines for conducting and reporting experiments or case studies \cite{Runeson_2009} to regulate their future evaluation work.

\section{Conclusions and future work}
\label{VIII}
Evaluation of commercial Cloud services has gradually become significant as an increasing number of competing Cloud providers emerge in industry \cite{Prodan_Ostermann_2009}\citeS{Li_Yang_2010}. Given that the Cloud services evaluation is challenging and the existing studies are relatively chaotic, we adopted the SLR method to investigate the existing practices as evidence to outline the scope of Cloud services evaluation. The findings of this SLR lie in three aspects. 
\begin{enumerate*}
\renewcommand{\labelenumi}{\it{(\theenumi)}}
    \item	\textbf{The overall data collected in the SLR can lead us to become familiar with the sate-of-the-practice of evaluation of commercial Cloud services.} In particular, the answers to those six research questions summarized the key details of the current evaluation implementations. Meanwhile, the summarized data, such as metrics, benchmarks, and experimental setup scenes, were arranged as a dictionary-like fashion for evaluators to facilitate future Cloud services evaluation work.
    \item	\textbf{Some of the findings have identified several research gaps in the area of Cloud services evaluation.} First, although Elasticity and Security are significant features of commercial Cloud services, there seems a lack of effective and efficient means of evaluating the elasticity and security of a Cloud service. Our findings also suggest that this could be a long-term research challenge. Second, there is still a gap between practice and research into ``real" Cloud evaluation benchmarks. On one hand, theoretical discussions considered that traditional benchmarks were insufficient for evaluating commercial Cloud services \cite{Binnig_Kossmann_2009}. On the other hand, traditional benchmarks have been overwhelmingly used in the existing Cloud evaluation practices. The findings suggest that those traditional benchmarks will remain in the Cloud services evaluation work unless there is a dedicated Cloud benchmark. Third, the result of a quality assessment of the studies shows that the existing primary studies were not always conducted or reported appropriately. Thus, we suggest that future evaluation work should be regulated following particular guidelines \cite{Budgen_Kitchenham_2008,Runeson_2009}.
    \item	\textbf{Some other findings suggest the trend of applying commercial Cloud services.} In general, commercial Cloud Computing has attracted the attention of an increasing number of researchers, which can be confirmed by the world-widely increased research interests in the Cloud services evaluation topic. In addition to satisfying business requirements, commercial Cloud Computing is also regarded as a suitable paradigm to deal with scientific issues. As for specific commercial Cloud services, although the competitive market changes rapidly, Amazon, Google and Microsoft currently supply the most popular Cloud services. Furthermore, PaaS and IaaS essentially supplement each other to satisfy various requirements in the Cloud market.
\end{enumerate*} 

We also gained some lessons about conducting SLR from this work. Firstly, our practice has confirmed some previous experiences like the usage of pilot review from other SLR studies \cite{Major_Kyriacou_2011,Barbar_Zhang_2009}. In particular, future studies should carefully estimate the extra time and effort if considering an ad hoc review as the prerequisite of an SLR conduction. Secondly, our study also revealed new EBSE lesson -- continuous collection of evidence for building knowledge base. In other words, for particular research topics, the employment of SLR could be worthy of a regular use to update the data or knowledge to support the research in those topics. In fact, given the initial understanding of Cloud services evaluation in this case, the current stage of this SLR tends to be a systematic mapping study, while the gradual update will accumulate the evaluation outcomes of more primary studies, and then help gain more knowledge. 

Our future work will be unfolded in two directions. Firstly, the extracted data in this SLR will be structured and stored into a database for supporting a Cloud services evaluation methodology. Secondly, benefiting from the result of this SLR as a solid starting point, we will perform deeper study into Cloud service evaluation, such as developing sophisticated evaluation metrics.

\section*{Acknowledgements}
We record our
sincere thanks for Prof. Barbara Kitchenham's pertinent suggestions and
comments that helped us improve the quality of this report.

NICTA is funded by the Australian Government as represented by the Department of Broadband, Communications and the Digital Economy and the Australian Research Council through the ICT Centre of Excellence program.

\bibliographystyle{model1b-num-names}
\setbiblabelwidth{99}
\bibliography{Ref}

\appendix

\section{Details of quality rating for primary studies}
See Table \ref{tbl>quality}.

\begin{table*}[!t]\footnotesize
\renewcommand{\arraystretch}{1.3}
\centering
\caption{\label{tbl>quality}Detailed score card for the quality assessment of the 82 primary studies.}

\begin{tabular}{l l l l l l l l l l l l}
\hline

\hline
Study & QA1 & QA2 & QA3 & QA4 & Research Reporting Score & QA5 & QA6 & QA7 & QA8 & Evaluation Reporting Score & Total Score\\
\hline

\citeS{Alhamad_Dillon_2010} & 0 & 1 & 1 &1 & 3 & 0.5 & 1 & 0 & 1 & 2.5 & 5.5\\
\citeS{Akioka_Muraoka_2010} & 1 & 1 & 0 & 1 & 3 & 0 & 1 & 0 & 1 & 2 & 5\\
\citeS{Bicer_Chiu_2011} & 1 & 1 & 1 & 1 & 4 & 0 & 1 & 1 & 1 & 3 & 7\\
\citeS{Brantner_Florescu_2008} & 1 & 1 & 1 & 1 & 4 & 0.5 & 1 & 1 & 1 & 3.5 & 7.5\\
\citeS{Bientinesi_Iakymchuk_2010} & 1 & 1 & 1 & 1 & 4 & 0.5 & 1 & 1 & 1 & 3.5 & 7.5\\
\citeS{Baun_Kunze_2009} & 1 & 1 & 0 & 1 & 3 & 0.5 & 1 & 0 & 0.5 & 2 & 5\\
\citeS{Brebner_Liu_2010} & 1 & 1 & 0 & 1 & 3 & 0.5 & 1 & 0 & 1 & 2.5 & 5.5\\
\citeS{Barker_Shenoy_2010} & 1 & 1 & 1 & 1 & 4 & 0.5 & 1 & 1 & 1 & 3.5 & 7.5\\
\citeS{Bermbach_Tai_2011} & 1 & 1 & 1 & 1 & 4 & 1 & 1 & 1 & 1 & 4 & 8\\
\citeS{Chiu_Agrawal_2010} & 1 & 1 & 1 & 1 & 4 & 0 & 1 & 1 & 1 & 3 & 7\\
\citeS{Chen_Bai_2011} & 1 & 1 & 0 & 1 & 3 & 0 & 1 & 1 & 1 & 3 & 6\\
\citeS{Carlyle_Harrell_2010} & 1 & 1 & 1 & 1 & 4 & 1 & 0.5 & 0.5 & 1 & 3 & 7\\
\citeS{Chiu_Hall_2011} & 1 & 1 & 1 & 1 & 4 & 0 & 1 & 1 & 1 & 3 & 7\\
\citeS{Cunha_Mendonca_2011} & 1 & 1 & 1 & 1 & 4 & 0 & 1 & 1 & 1 & 3 & 7\\
\citeS{Cervino_Rodriguez_2011} & 1 & 1 & 1 & 1 & 4 & 0 & 1 & 1 & 1 & 3 & 7\\
\citeS{Assuncao_Costanzo_2010} & 1 & 1 & 1 & 1 & 4 & 0 & 1 & 1 & 1 & 3 & 7\\
\citeS{Dalman_Doernemann_2010} & 1 & 1 & 1 & 1 & 4 & 0 & 1 & 1 & 1 & 3 & 7\\
\citeS{Dejun_Pierre_2010} & 1 & 1 & 1 & 1 & 4 & 0.5 & 1 & 1 & 1 & 3.5 & 7.5\\
\citeS{Deelman_Singh_2008} & 1 & 1 & 1 & 1 & 4 & 0 & 1 & 1 & 1 & 3 & 7\\
\citeS{Evangelinos_Hill_2008} & 1 & 1 & 0 & 1 & 3 & 0 & 0.5 & 0.5 & 1 & 2 & 5\\
\citeS{Gropengie_Baumann_2011} & 1 & 1 & 1 & 1 & 4 & 0 & 0.5 & 1 & 1 & 2.5 & 6.5\\
\citeS{Ghoshal_Canon_2011}  & 1 & 1 & 1 & 1 & 4 & 1 & 1 & 1 & 1 & 4 & 8\\
\citeS{Garfinkel_Journal_2007} & 0 & 1 & 0 & 1 & 2 &0 & 0.5 & 0.5 & 1 & 2 & 4\\
\citeS{Gent_Kotthoff_2011} & 1 & 1 & 1 & 1 & 4 & 0 & 1 & 0.5 & 1 & 2.5 & 6.5\\
\citeS{Garfinkel_Report_2007} & 1 & 1 & 1 & 1 & 4 & 1 & 1 & 1 & 1 & 4 & 8\\
\citeS{Gropengie_Sattler_2011} & 1 & 1 & 1 & 1 & 4 & 0 & 1 & 0.5 & 1 & 2.5 & 6.5\\
\citeS{Gunarathne_Wu_2011} & 0 & 1 & 1 &1 & 3 & 0 & 1 & 1 &1 & 3 & 6\\ 
\citeS{Hazelhurst_2008} & 1 & 1 & 0 & 1 & 3 & 0 & 1 & 1 & 1 & 3 & 6\\
\citeS{Hill_Humphrey_2009} & 1 & 1 & 1 & 1 & 4 & 0 & 1 & 1 & 1 & 3 & 7\\
\citeS{Humphrey_Hill_2011} & 1 & 1 & 1 & 1 & 4 & 1 & 1 & 1 & 1 & 4 & 8\\
\citeS{Hill_Li_2010} & 1 & 1 & 1 & 1 & 4 & 1 & 1 & 1 & 1 & 4 & 8\\
\citeS{He_Zhou_2010} & 1 & 1 & 1 & 1 & 4 & 0 & 1 & 0.5 & 1 & 2.5 & 6.5\\
\citeS{Islam_Lee_2011} & 1 & 1 & 1 & 1 & 4 & 0 & 1 & 1 & 1 & 3 & 7\\
\citeS{Iakymchuk_Napper_2011} & 1 & 1 & 0 & 1 & 3 & 0 & 1 & 1 & 1 & 3 & 6\\
\citeS{Iosup_Ostermann_2011}& 1 & 1 & 1 & 1 & 4 & 0 & 1 & 1 & 1 & 3 & 7\\
\citeS{Iosup_Yigitbasi_2011} & 1 & 1 & 1 & 1 & 4 & 1 & 1 & 1 & 1 & 4 & 8\\
\citeS{Juve_Deelman_2009} & 1 & 1 & 0.5 & 1 & 3.5 & 0 & 1 & 1 & 1 & 3 & 6.5\\
\citeS{Juve_Deelman_2010} & 1 & 1 & 1 & 1 & 4 & 0 & 1 & 1 & 1 & 3 & 7\\
\citeS{Juve_Deelman_2011} & 1 & 1 & 0.5 & 1 & 3.5 & 0 & 1 & 1 & 1 & 3 & 6.5\\
\citeS{Jackson_Muriki_2011} & 1 & 1 & 1 & 1 & 4 & 0.5 & 1 & 1 & 1 & 3 & 7.5\\
\citeS{Jayasinghe_Malkowski_2011} & 1 & 1 & 1 & 1 & 4 & 0 & 1 & 1 & 1 & 3 & 7\\
\citeS{Jackson_Ramakrishnan_2010} & 1 & 1 & 1 & 1 & 4 & 0 & 1 & 1 & 1 & 3 & 7\\
\citeS{Kondo_Javadi_2009} & 1 & 1 & 1 & 1 & 4 & 0 & 1 & 0.5 & 1 & 2.5 & 6.5\\
\citeS{Khamra_Kim_2010} & 1 & 1 & 1 & 1 & 4 & 0 & 0.5 & 0.5 & 1 & 2 & 6\\
\citeS{Kossmann_Kraska_2010} & 1 & 1 & 1 & 1 & 4 & 0.5 & 1 & 1 & 1 & 3.5 & 7.5\\
\citeS{Li_Humphrey_2010} & 1 & 1 & 1 & 1 & 4 & 0 & 1 & 1 & 1 & 3 & 7\\
\citeS{Luckow_Jha_2010} & 1 & 1 & 0 & 1 & 3 & 0.5 & 1 & 0.5 & 1 & 3 & 6\\
\citeS{Lu_Jackson_2010} & 1 & 1 & 1 & 1 & 4 & 0 & 1 & 1 & 1 & 3 & 7\\
\citeS{Lenk_Menzel_2011} & 1 & 1 & 1 & 1 & 4 & 1 & 1 & 1 & 1 & 4 & 8\\ 
\citeS{Liu_Wee_2009} & 1 & 1 & 0 & 1 & 3 & 0.5 & 1 & 1 & 1 & 3.5 & 6.5\\
\citeS{Li_Yang_2010} & 1 & 1 & 1 & 1 & 4 & 1 & 0.5 & 0.5 & 1 & 3 & 7\\

\multicolumn{12}{r}{Continued on next page} \\
\end{tabular}

\end{table*}

\begin{table*}[!t]\footnotesize
\renewcommand{\arraystretch}{1.3}
\centering
\begin{tabular}{l l l l l l l l l l l l}

\multicolumn{12}{l}{Table \ref{tbl>quality} (continued from previous page)} \\
\hline

\hline
Study & QA1 & QA2 & QA3 & QA4 & Research Reporting Score & QA5 & QA6 & QA7 & QA8 & Evaluation Reporting Score & Total Score\\
\hline
\citeS{Liu_Zhai_2011} & 1 & 1 & 1 & 1 & 4 & 0 & 1 & 0 & 1 & 2 & 6\\
\citeS{Montella_Foster_2010} & 0.5 & 1 & 0 & 1 & 2.5 & 0 & 1 & 1 & 1 & 3 & 5.5\\
\citeS{Napper_Bientinesi_2009} & 1 & 1 & 0.5 & 1 & 3.5 & 0.5 & 1 & 1 & 1 & 3.5 & 7\\
\citeS{Ostermann_Iosup_2009} & 1 & 1 & 1 & 1 & 4 & 0.5 & 1 & 1 & 1 & 3.5 & 7.5\\
\citeS{Phillips_Engen_2011} & 1 & 1 & 1 & 1 & 4 & 0 & 1 & 1 & 1 & 3 & 7\\
\citeS{Palankar_Iamnitchi_2008} & 1 & 1 & 0 & 1 & 3 & 1 & 1 & 1 & 1 & 4 & 7\\
\citeS{Ramasahayam_Deters_2011} & 1 & 1 & 0 & 1 & 3 & 0 & 0.5 & 0.5 & 1 & 2 & 5\\
\citeS{Redekopp_Simmhan_2011} & 1 & 1 & 1 & 1 & 4 & 0 & 1 & 1 & 1 & 3 & 7\\
\citeS{Ristenpart_Tromer_2009} & 1 & 1 & 0 & 1 & 3& 0 & 1 & 1 & 1 & 3 & 6\\
\citeS{Rehr_Vila_2010} & 1 & 1 & 0 & 1 & 3 & 0 & 1 & 0.5 & 1 & 2.5 & 5.5\\
\citeS{Salah_Saba_2011} & 1 & 1 & 1 & 1 & 4 & 0 & 1 & 1 & 1 & 3 & 7\\
\citeS{Schad_Dittrich_2010} & 1 & 1 & 1 & 1 & 4 & 1 & 1 & 1 & 1 & 4 & 8\\
\citeS{Schatz_Koschnicke_2011} & 0.5 & 1 & 0.5 & 1 & 3 & 0.5 & 1 & 1 & 1 & 3.5 & 6.5\\
\citeS{Subramanian_Ma_2011} & 0 & 1 & 1 & 1 & 3 & 0 & 0.5 & 1 & 1 & 2.5 & 5.5\\
\citeS{Sobel_Subramanyam_2008} & 1 & 1 & 0 & 1 & 3 & 0 & 1 & 0.5 & 1 & 2.5 & 5.5\\
\citeS{Stantchev_2009} & 1 & 1 & 1 & 1 & 4 & 0.5 & 1 & 1 & 1 & 3.5 & 7.5\\
\citeS{Tran_Cinquini_2011} & 1 & 1 & 1 & 1 & 4 & 0.5 & 1 & 1 & 1 & 3.5 & 7.5\\
\citeS{Toyoshima_Yamaguchi_2010} & 0 & 1 & 0 & 1 & 2 & 0 & 0.5 & 0 & 0.5 & 1 & 3\\
\citeS{Virvilis_Dritsas_2011} & 1 & 1 & 0 & 1 & 3 & 0 & 0.5 & 1 & 0.5 & 2 & 5\\
\citeS{Vockler_Juve_2011} & 1 & 1 & 1 & 1 & 4 & 0 & 1 & 1 & 1 & 3 & 7\\
\citeS{Vozmediano_Montero_2011}& 1 & 1 & 1 & 1 & 4 & 0 & 1 & 1 & 1 & 3 & 7\\
\citeS{Vecchiola_Pandey_2009} & 1 & 1 & 0 & 1 & 3 & 0 & 1 & 0.5 & 1 & 2.5 & 5.5\\
\citeS{Walker_2008} & 1 & 1 & 0 & 1 & 3 & 0 & 1 & 1 & 1 & 3 & 6\\
\citeS{Wall_Kudtarkar_2010} & 1 & 1 & 1 & 1 & 4 & 0 & 1 & 0.5 & 1 & 2.5 & 6.5\\
\citeS{Wang_Ng_2010} & 1 & 1 & 1 & 1 & 4 & 0.5 & 1 & 1 & 1 & 3.5 & 7.5\\
\citeS{Wang_Varman_2011} & 1 & 1 & 1 & 1 & 4 & 0 & 1 & 1 & 1 & 3 & 7\\
\citeS{Wilkening_Wilke_2009} & 1 & 1 & 0 & 1 & 3 & 0.5 & 1 & 0.5 & 1 & 3 & 6\\
\citeS{Yigitbasi_Iosup_2009} & 1 & 1 & 1 & 1 & 4 & 0 & 0.5 & 1 & 1 & 2.5 & 6.5\\
\citeS{Zaspel_Griebel_2011} & 0.5 & 1 & 0 & 1 & 2.5 & 0 & 1 & 1 & 1 & 3 & 5.5\\
\citeS{Zhao_Liu_2010} & 1 & 1 & 1 & 1 & 4 & 1 & 0.5 & 1 & 1 & 3.5 & 7.5\\
\citeS{Zhai_Liu_2011}& 1 & 1 & 1 & 1 & 4 & 1 & 1 & 1 & 1 & 4 & 8\\
\hline
Total & 75.5 & 82 & 58 & 82 & 297.5 & 22 & 76 & 68 & 80.5 & 246 & 544\\
Average & 0.92 & 1 & 0.71 & 1 & 3.63 & 0.27 & 0.93 & 0.83 & 0.98 & 3 & 6.63\\
\hline

\hline
\end{tabular}

\end{table*}

\section{Brief description of the evaluated commercial Cloud services}
\label{Appendix>II}
See Table \ref{tbl>services}.
\begin{table*}[!t]\footnotesize
\renewcommand{\arraystretch}{1.3}
\centering
\caption{\label{tbl>services}Evaluated commercial Cloud services.}
\begin{tabular}{p{1.5cm} l >{\raggedright\arraybackslash}p{11cm}}
\hline

\hline
Cloud Provider & Cloud Service & Brief Description\\
\hline
\multirow{16}{*}{Amazon} & EBS (Elastic Block Store) & Amazon Elastic Block Store (EBS) provides block level storage volumes for use with Amazon EC2 instances.\\
 & EC2 (Elastic Compute Cloud) & Amazon Elastic Compute Cloud (Amazon EC2) provides resizable compute capacity in the cloud.\\
 & ELB (Elastic Load Balancing) & Elastic Load Balancing automatically distributes incoming application traffic across multiple Amazon EC2 instances.\\
 & EMR (Elastic MapReduce) & Amazon Elastic MapReduce enables businesses, researchers, data analysts, and developers to easily and cost-effectively process vast amounts of data.\\
 & FPS (Flexible Payment Service) & Amazon FPS is built on top of Amazon's payments infrastructure and provides developers with a convenient way to charge Amazon's tens of millions of customers.\\
 & RDS (Rational Database Service) & Amazon Relational Database Service (Amazon RDS) is used to set up, operate, and scale a relational database in the cloud.\\
 & S3 (Simple Storage Service) & Amazon S3 provides a simple web services interface that can be used to store and retrieve any amount of data, at any time, from anywhere on the web.\\
 & SimpleDB & Amazon SimpleDB is a non-relational data store that offloads the work of database administration.\\
 & SQS (Simple Queueing System) & Amazon Simple Queue Service (Amazon SQS) offers a hosted queue for storing messages as they travel between computers.\\
\cline{2-3}
\multirow{2}{*}{BlueLock} & BlueLock & Bluelock Virtual Datacenters are hosted in the public cloud and are based on VMware vCloud technology, which provides full compatibility with any VMware environment.\\
\cline{2-3}
\multirow{2}{*}{ElasticHosts} & ElasticHosts & ElasticHosts supplies virtual servers running on server farms, located in five fully-independent premier-class data centres across two continents.\\
\cline{2-3}
\multirow{2}{*}{Flexiant} & FlexiScale & Flexible \& Scalable Public Cloud Hosting is a pay-as-you-go public cloud platform offering on-demand, scalable hosting services.\\
\cline{2-3}
\multirow{2}{*}{GoGrid} & GoGrid & GoGrid is a cloud infrastructure service, hosting Linux and Windows virtual machines managed by a multi-server control panel.\\
\cline{2-3}
\multirow{6}{*}{Google} & AppEngine (Google App Engine) & Google AppEngine is a cloud computing platform for developing and hosting web applications in Google-managed data centres.\\
 & Memcache & Memcache is a distributed memory object caching system, primarily intended for fast access to cached results of datastore queries.\\
 & UrlFetch (URL Fetch) & UrlFetch allows scripts to communicate with other applications or access other resources on the web by fetching URLs.\\
\cline{2-3}
IBM & IBM Cloud (Beta) & The beta version of Cloud computing platform offered by IBM.\\
\cline{2-3}
\multirow{4}{*}{Microsoft} & SQL Azure & Microsoft SQL Azure Database is a cloud database service built on SQL Server technologies.\\
 & Windows Azure & Windows Azure is a cloud operating system that serves as a runtime for the applications and provides a set of services that allows development, management and hosting of applications off-premises.\\
\cline{2-3}
\multirow{4}{*}{Rackspace} & CloudServers & CloudServers is a cloud infrastructure service that allows users to deploy \textquotedblleft one to hundreds of cloud servers instantly\textquotedblright { }and create of \textquotedblleft advanced, high availability architectures\textquotedblright .\\
 & CloudFiles & CloudFiles is a cloud storage service that provides \textquotedblleft unlimited online storage and CDN\textquotedblright { }for media on a utility computing basis.\\
\hline

\hline
\end{tabular}
\end{table*}

\section{Explanation of the typically excluded papers}
\label{Appendix>III}
See Table \ref{tbl>3aa}. We only show typical publications here instead of listing all the excluded studies. Most of the typically excluded papers were discussed in our group meetings. This appendix may be used as a clue for readers to further identify useful information.

\begin{table}[!t]\footnotesize
\renewcommand{\arraystretch}{1.3}
\centering
\caption{\label{tbl>3aa}Explanation of the typically excluded papers.}
\begin{tabular}{p{1.1cm} >{\raggedright}p{4.9cm} p{1.5cm}}
\hline

\hline
Paper & Brief Explanation & Corresponding Exclusion Criteria\\
\hline

\citeX{Bunch_Chohan_2010} & The evaluation work is for the proposed AppScale Cloud platform. & (3)\\
\citeX{Binnig_Kossmann_2009} & Theoretical discussion about Cloud services evaluation. & (2)\\
\citeX{Bruneo_Longo_2011} & The evaluation work is for the proposed modeling approach, and it is in a private virtualized environment. & (1) \& (3)\\
\citeX{Cooper_Silberstein_2010} & Mostly theoretical discussion, and evaluation work is in a private environment. & (1) \& (2)\\
\citeX{Assuncao_Costanzo_2009} & This is a previous version of \citeS{Assuncao_Costanzo_2010}. & (4)\\
\citeX{Ekanayake_Fox_2009} & The evaluation work is done in the open-source Cloud. & (1)\\
\citeX{Edlund_Koopmans_2011} & Theoratical discussion based on the evaluation work in a private Cloud. & (1) \& (2)\\ 
\citeX{Gao_Lowe_2009} & The evaluation work is for the proposed VBS system. & (3)\\
\citeX{Gupta_Milojicic_2011} & The evaluation work is done in the open-source Cloud. & (1)\\
\citeX{Graubner_Schmidt_2011}& The evaluation work is done in the open-source Cloud. & (1)\\
\citeX{Gunarathne_Wu_2010a} & This is a previous version of \citeS{Gunarathne_Wu_2011}. & (4)\\
\citeX{Gunarathne_Wu_2010} & The evaluation work is for the proposed AzureMapReduce framework. & (3)\\
\citeX{Haak_Menzel_2011} & Theoretical discussion about autonomic benchmarking Cloud services. & (2)\\
\citeX{Huber_Quast_2011} & The evaluation work is done in a private virtualized environment. & (1)\\
\citeX{Iosup_Yigitbasi_2010_excluded} & This is a previous version of \citeS{Iosup_Yigitbasi_2011}. & (4)\\
\citeX{Jackson_Ramakrishnan_2010_excluded} & This is a previous version of \citeS{Jackson_Muriki_2011}. & (4)\\
\citeX{Kaur_Chana_2011} & This is a poster paper. & (5)\\
\citeX{Kobayashi_Mikami_2011} & The evaluation work is in a private Cloud. & (1)\\
\citeX{Liu_Orban_2008} & This work is for the proposed GridBatch with little evaluation. & (3)\\
\citeX{Ostermann_Iosup_2008_excluded} & This is a previous version of \citeS{Ostermann_Iosup_2009}. & (4)\\
\citeX{Ostermann_Prodan_2009} & The evaluation work is in a private Cloud. & (1)\\
\citeX{Pallickara_Pierce_2009} & The evaluation work is for the proposed Swarm framework. & (3)\\
\citeX{Rehman_Sakr_2010} & The evaluation work is done in an academic Cloud: Qloud. & (1)\\
\citeX{Schatz_2009} & The evaluation work is for the proposed MapReduce-based algorithm. & (3)\\
\citeX{Shafer_2010} & The evaluation work is done in the open-source Cloud. & (1)\\
\citeX{Sivathanu_Liu_2010} & The evaluation work is done in a private virtualized environment. & (1)\\
\citeX{Turcu_Foster_2011} & The evaluation work is for the proposed scheduling strategy. & (3)\\
\citeX{Tak_Urgaonkar_2011} & The evaluation work is done in a private virtualized environment. & (1)\\
\citeX{Voorsluys_Brokerg_2009} & The evaluation work is not on commercial Cloud services. & (1)\\
\citeX{Wang_Varman_2010} & This is a previous version of \citeS{Wang_Varman_2011}. & (4)\\
\citeX{Yang_Tan_2009} & Mainly a theoratical discussion about performance evaluation with fault recovery. & (2)\\
\hline

\hline
\end{tabular}
\end{table}

\bibliographystyleS{alpha}
\bibliographyS{SLR_Selected}

\bibliographystyleX{alpha}
\bibliographyX{SLR_Excluded}

\end{document}